\documentclass[aps,prb,reprint,showpacs]{revtex4-1}

\usepackage{graphicx}
\usepackage{dcolumn}
\usepackage{bm}
\usepackage{epsfig}
\usepackage{amsmath}
\usepackage[usenames,dvipsnames]{color}
\usepackage{booktabs}
\begin{document}

\author{Mahmoud M. Asmar}
\email{asmar@phy.ohiou.edu} \affiliation{Department of Physics and
Astronomy and Nanoscale and Quantum Phenomena Institute, Ohio
University, Athens, Ohio 45701-2979}
\author{Sergio E. Ulloa}
\email{ulloa@ohio.edu} \affiliation{Department of Physics and
Astronomy and Nanoscale and Quantum Phenomena Institute, Ohio
University, Athens, Ohio 45701-2979}

\title{Symmetry breaking effects on spin and electronic transport in graphene}

\begin{abstract}
The decoration of graphene samples with adatoms or nanoparticles leads to the
enhancement of spin-orbit interactions as well as to the
introduction of symmetry breaking effects that could have drastic
effects on spin and electronic transport phenomena. We present an analysis based on
symmetry considerations and examine the impact on the scattering
matrix for graphene systems containing defects that enhance
spin-orbit interactions, while conserving the electronic total
angular momentum.  We show that the appearance and
dominance of skew scattering, and  the related observation of valley and/or spin
Hall effect in decorated graphene samples, suggests the set of symmetries that
adatom perturbations should satisfy.
We further show that detailed measurements of
the transport and elastic times as function of carrier concentration make it possible to
not only extract the strength of the spin-orbit interaction, as suggested before, but also
obtain the amplitude of the symmetry breaking terms introduced.  To examine how
different terms would affect measurements, we also present calculations for typical
random distributions of impurities with different perturbations, illustrating the detailed
energy dependence of different observables.
\end{abstract}


\maketitle

\section{Introduction.}
The low energy Dirac fermions residing near the corners of the
Brillouin zone on the honeycomb lattice of
graphene~\cite{elctronincproperties}  have inspired the theoretical
prediction and subsequent experimental realization of a variety of
new two-dimensional materials~\cite{quntums,timerev} and novel
electronic phases of matter in two and three
dimensions.~\cite{topoexp1,topol3,topol2,topol1} The ability to
manipulate and control the electronic and spin properties of
graphene and related materials by decoration,
\cite{impurityso,colossal,ferreira1,nanoparticles} stacking,
\cite{layered,layered1}
intercalation\cite{Ni1,golddep,Nii,rossi,moire} and distortion
\cite{curvAndo,newcurv1,newcurv2,curvature2,cham1,keku,curvature3,cham3,cham2} may also lead to
important technological advances.

The helical nature of the massless Dirac particles in pristine
graphene gives rise to interesting phenomena, such as Klein
tunneling,~\cite{Klein} and the observation of weak
anti-localization in this material.~\cite{disord8} However, it has
been shown that the transport properties of graphene strongly depend
on the nature, symmetries and concentration of the impurities
present in typical samples.
\cite{disord1,disord2,disord4,disord6,disord3,disord8,scatteringtimes,helen,disord7,disord5,nonabelian,
ferreira2} Deformations from impurities may belong to different
symmetry classes.  As such, they could either enhance or weaken the
signatures expected from the chiral nature of the scatterers of
graphene and drastically modify the electronic and spin transport
properties.

An interesting example of interactions protected by lattice symmetries
is provided by the appearance of a spin orbit interaction (SOI) that
preserves all symmetries of the honeycomb lattice,~\cite{quntums}
and leads to the formation of topologically protected edge
states--anticipating the existence of topological insulating
materials. \cite{timerev}  In contrast, an electric field
perpendicular to a graphene sample leads to inversion ($z
\rightarrow -z$) symmetry breaking, allowing the appearance of a
Bychkov-Rashba SOI.~\cite{quntums,Rashaba} The small atomic number
of carbon makes an intrinsic SOI small (a few $\mu$eV),\cite{intrinsic}
while the Rashba SOI has strength of hundreds of  $\mu$eV
for typical values of electric fields.~\cite{intrinsic}

Methods that rely on the artificial functionalization of graphene
could lead to sizable enhancement of SOIs. Successful
examples of these methods include the intercalation of gold atoms
between graphene and the Ni substrate,~\cite{Ni1,golddep,Nii} weak
hydrogenation,~\cite{colossal,impurityso} and the deposition of
large copper or gold nanoparticles. \cite{nanoparticles} Other
approaches are also promising, including externally induced
curvature of samples,~\cite{curvAndo,curvature2,curvature3} decoration with
heavy adatoms such as thallium and indium,~\cite{indium,indiumexpt}  and the
deposition of graphene on the surface of a strong topological
material.~\cite{rossi} These methods have achieved or project the
enhancement of SOI of up to several tens of meV, leading to clear
measurable effects on the electron
dynamics.~\cite{isotropic,ferreira1,ferreira2,impurityso,mediateso,ellyot,biref}

The methods used to enhance SOIs in graphene require the
presence of adatoms or local defects/deformations that accommodate
on the graphene lattice in different ways.  These perturbations
introduce additional effects from the lattice symmetries they may
break. The enhancement of intrinsic and/or Rashba SOIs on the
dynamics of electrons in graphene has been analyzed for the most
part in the framework of a single-valley Dirac description.
\cite{cluster,isotropic,impurityso,ferreira1} In that limit--valid
for potential scatterers with large characteristic spatial
variation compared to the lattice constant--the scattering preserves
helicity, it is predominantly forward, and gives rise to
Klein-tunneling, as mentioned
above.~\cite{helen,scatteringtimes,isotropic} If the scattering
centers lead to enhancement of Rashba and/or intrinsic SOIs, the
electronic scattering has been shown to become increasingly
isotropic.~\cite{isotropic} This scattering could also lead to spin
Hall effect via enhanced skew scattering close to the resonant
regime.\cite{ferreira1,colossal,nanoparticles}

Here, going beyond the single-valley Dirac treatment,\cite{isotropic} we study the scattering properties of Dirac fermions from
adatoms or imperfections on graphene, as they may lead to the reduction
of lattice symmetries. Through a symmetry analysis and considering
intra- and intervalley processes, we are able to identify the
constants of motion for different situations.  We further examine the
constraints imposed by these conserved quantities on the scattering
matrix and the observable consequences in transport experiments.

We focus in particular on the ratio of transport to elastic times, $\xi=\tau_{tr}/\tau_e$, and its
dependence on carrier concentration, which was previously studied for systems where SOIs were locally enhanced,
and no additional local symmetry breaking effects were considered. \cite{isotropic}  We find
that this ratio depends strongly on the microscopic potential perturbations,
which make $\xi$ deviate from its characteristic low-energy value of $2$. \cite{isotropic}
This deviation depends on
the carrier concentration and strength of the symmetry breaking, as
well as on the local potential shift produced by the impurities.
As backscattering becomes stronger, the ratio takes a
value closer to $1$ (the limit for massive Schr\"odinger particles).
We further study the effects symmetry
breaking perturbations may produce on the detection of Rashba and intrinsic SOIs.
The dependence of $\xi$ on carrier
concentration in the presence of Rashba SOI is shown to be
{\em qualitatively} different for
different symmetry breaking terms in the perturbation. Systematic
experimental studies of $\xi$ in given graphene systems would allow the
identification and simultaneous evaluation of the relative strengths of the
different perturbations.
Finally, from the symmetries of the scattering matrix, we show
which systems may display non-zero valley and/or spin transport skewness (skew scattering).  In particular,
we find that disorder that is not invariant under ``effective time
reversal'' leads to a reduction of spin skew scattering, due to the enhancement of valley skew scattering; this allows one to
identify what kind of decorated system will display
skew scattering and may result in the appearance of valley and/or spin Hall effect.

We further illustrate the impact of the symmetry considerations described above for typical
graphene samples.  Throughout the discussion of the different perturbations, we
present numerical results for the anticipated dependence of $\xi$ and skewness to
different effects, by analyzing the results for a random impurity distribution.  This allows
us to gain insights into the contribution of different effects to various measurable
quantitites.

The remainder  of the paper considers the symmetries of different perturbation
terms, especially their invariance under time reversal, and the consequences of
these symmetries on the corresponding scattering matrix in the problem.  In  Sec.\
\ref{sect:spinless} we consider the spinless case, and study the scattering matrix
and scattering amplitudes, as well as their impact in transport and elastic times.
In Sec.\ \ref{sect:spin} we include the spin degree of freedom and associated
spin-orbit interaction terms, while Sec.\ \ref{sect:skew} studies the consequences
on skew scattering.  A general discussion of the results in Sec.\ \ref{sect:discussion}
is followed by conclusions in Sec.\ \ref{sect:conclude}.

\section{Symmetry of perturbations in graphene}\label{sect:model}
\subsection{Spinless scattering} \label{sect:spinless}

In order to analyze the effects of symmetry breaking on the dynamics
of Dirac fermions in graphene, we first consider the Hamiltonian
\begin{equation}
H=H_{0}+V_{T}\;,
\end{equation}
where
\begin{equation}
H_{0}=v_{F}\vec{\alpha} \cdot \vec{k}\; ,
\end{equation}
($\hbar=1$ and $\vec{k}=(k_{x},k_{y})$) is the pristine graphene
Hamiltonian, represented in the four-dimensional chiral basis
$\psi=\left(\psi_{A,K},\psi_{B,K},\psi_{B,K'},\psi_{A,K'}\right)^{T}$,
with
\begin{equation}
\alpha_{i}=\tau_{3}\otimes\sigma_{i}=\left(
                                       \begin{array}{cc}
                                         \sigma_{i} & 0\\
                                         0 & -\sigma_{i} \\
                                       \end{array}
                                     \right)\;,
\end{equation}
$(i=1,2,3)$, where $\sigma_{i}$ and
$\tau_{i}$ are a set of Pauli matrices that act on the sublattice
index ($A,B$) and the valley index ($K,K'$), respectively.
It is also useful to define
\begin{eqnarray}
I=\sigma_{0}\otimes\tau_{0}=\left(
                              \begin{array}{cc}
                                \sigma_{0} & 0 \\
                                0 & \sigma_{0} \\
                              \end{array}
                            \right)\;,\\
\gamma^{5}=\tau_{3}\otimes\sigma_{0}=\left(
                                       \begin{array}{cc}
                                         \sigma_{0} & 0 \\
                                         0 & -\sigma_{0} \\
                                       \end{array}
                                     \right)\;,
\end{eqnarray}
and
\begin{equation}
\beta=\tau_{1}\otimes \sigma_{0}=\left(
                                   \begin{array}{cc}
                                     0 & \tau_{0} \\
                                     \tau_{0} & 0 \\
                                   \end{array}
                                 \right)\; .
\end{equation}

As in the
case of the three dimensional Dirac equation,~\cite{peskin,book}
these matrices satisfy
\begin{eqnarray}
\{\alpha_{i},\alpha_{j}\}&=&2\delta_{i,j}\;,\\
\left[\alpha_{i},\gamma^{5}\right]&=&0\;,\\
\{\alpha_{i},\beta\}&=&0\;,
\end{eqnarray}
and
\begin{equation}
\{\gamma^{5},\beta\}=0\;.
\end{equation}

\subsubsection{Symmetry considerations.}
Here we define, $V_{T}$, as the set of time reversal
invariant perturbations,~\cite{disord1,disord4,disord6,disord7}
\begin{equation}
[V_{T},\mathcal{T}]=0\;,
\end{equation}
where
\begin{equation}
\mathcal{T}=\beta\gamma^{5}\alpha_{1}\mathcal{C}\;
 \end{equation}
is the time reversal operator, $\mathcal{T}^2=1$, and $\mathcal{C}$ is the anti-linear complex
conjugation operator.~\cite{masses,BeenakkerRMP,isotropic}
Additionally, the helicity operator
\begin{equation}
\hat{h}=\gamma^{5}\vec{\alpha}\cdot
\vec{k}/k\;,
\end{equation}
(with $k=|\vec{k}|$) indicates the pseudo-spin projection in the direction
of momentum, \cite{masses,peskin} with
\begin{equation}
\left[H_{0},\hat{h}\right]=0\;,
\end{equation}
and
\begin{equation}
\{\mathcal{T},\hat{h}\}=0\;.
\end{equation}

The conservation of total angular momentum in a system in which
perturbations $V_{T}$ are present requires
\begin{equation}
\left[H,J_{z}\right]=0\;,
\end{equation}
where
\begin{equation}
J_{z}=-i\partial_{\theta}+\frac{1}{2}\gamma^{5}\alpha_{3} \;.
\end{equation}

In systems
where the spatial dependence of the perturbations has circular
symmetry, such as the simple form
$
H=H_{0}+V_{T}\Theta(R-r) ,
$ 
there exist a subset of time reversal invariant interactions that conserve
the total angular momentum,
\begin{equation}
V_{T,J}\subset V_{T}\;.
\end{equation}
This set can be shown to be composed of the elements~ \cite{masses,disord1,disord2,disord3,disord4,disord5,disord6,disord7,nonabelian,cham1,cham2,cham3}
\begin{equation}
V_{T,J}=\{v
I,s\alpha_{3},t\beta e^{i\gamma^{5}\phi}\}\;,
\end{equation}
where $v$, $s$, $t$ and $\phi$ are real parameters. Notice that with
the exception of  $v I$, the $V_{T,J}$ terms anticommute with
$\alpha_{1}$ and $\alpha_{2}$, so that they generate
gaps in the spectrum and are referred to as ``massive
terms''.~\cite{masses} The $v$ term describes a constant local
potential shift caused by the adatoms, while the $s$ term
describes a staggered potential effect some perturbations
may add, where the $A$ and $B$ sites are affected differently.
The $t$ terms describe a hopping modulation amplitude
caused by the adatoms/defects, where $\phi$ describes
the angle of dimerization of the resulting deformation, as in a
Kekul\'e pattern.~\cite{masses,cham1,cham2}

One can also define the subset
\begin{equation}
 V_{T,J,h}\subset V_{T,J}\;,
 \end{equation}
for perturbations that in addition to commuting with $\mathcal{T}$ and
$J_{z}$, also commute with the helicity operator $\hat{h}$, and find
that
\begin{equation}
V_{T,J,h}=\{v I, t \beta e^{i\gamma^{5}\phi}\}\;.
\end{equation}
In these cases the eigenstates of the Hamiltonian,
\begin{equation}
H=H_{0}+V_{T,J,h}\;,
\end{equation}
can be identified by corresponding quantum numbers, such that
\begin{equation}
\hat{h}\psi_{\pm,j}=\pm\psi_{\pm,j}
\end{equation}
(as $\hat{h}^2=1$), and
\begin{equation}
J_{z}\psi_{\pm,j}=j\psi_{\pm,j}\;,
\end{equation}
where $j$ is half-integer. One
can also find an additional operator
\begin{equation}
\mathcal{T}_{1}=ie^{-i\gamma^{5}\phi}\gamma^{5}\alpha_{2}\mathcal{C}\;,
\end{equation}
such that $\mathcal{T}^{2}_{1}=-1$,
\begin{equation}
[\mathcal{T}_{1},H]=[\mathcal{T}_{1},\hat{h}]=0\;,
\end{equation}
 while
 \begin{equation}
\{\mathcal{T}_{1},J_{z}\}=0\;.
\end{equation}
$\mathcal{T}_{1}$ can be seen as a
combination of a chiral rotation and the effective time reversal
operator in each Dirac
point.~\cite{masses,elastic,BeenakkerRMP,resonant2} This operator
together with the time reversal operator $\mathcal{T}$, relate the
states with angular momentum $j$ and $-j$, which can be chosen such
that
\begin{equation}
\mathcal{T}_{1}\psi_{\pm,j}=\pm(-1)^{j}\psi_{\pm,-j}\;,
\end{equation}
 and
 \begin{equation}
\mathcal{T}\psi_{\pm,j}=\mp(-1)^{j+\frac{1}{2}}\psi_{\mp,-j} \end{equation}
(see appendix~\ref{appenda}).

\subsubsection{Symmetry constraints on scattering and electronic transport.}

Conservation of total angular momentum allows one to use the
$j$ block-diagonal character of the scattering matrix. The lack of
valley mixing  in the area outside the potential perturbation region, $r>R$,
enables the determination of the scattering amplitudes between the
$+$ and $-$ (or correspondingly $K$ and $K'$) states for each $j$
block. The components of each $4\times4$ block ($\hat{S}_{j}$) are
given by
\begin{equation}
\hat{S}_{j,\tau\tau'}=S_{j,\tau\tau'} \sigma_{0}\;,
\end{equation}
where
$\tau,\tau'=\pm$. Unitarity of $\hat{S}$ requires
\begin{equation}
\hat{S}_{j}\hat{S}^{\dag}_{j}=I\;,
\end{equation}
and symmetry under operators $\mathcal{T}_{1}$
and $\mathcal{T}$ determine
relations between the $j$ and $-j$ blocks, such that
\begin{equation}
\gamma^{5}\mathcal{T}_{1}S_{j}\mathcal{T}^{-1}_{1}\gamma^{5}=S^{\dag}_{-j}
\end{equation}
and
\begin{equation}
\gamma^{5}\mathcal{T}S_{j}\mathcal{T}^{-1}\gamma^{5}=S^{\dag}_{-j}
\end{equation}
(see appendix~\ref{appenda}). These relations have
important consequences on the possible scattering processes (see
Eq.\ \ref{eq1a} in appendix), which allow us to obtain different scattering
functions.

For example, the differential cross section $\sigma(\theta)$ contains
information on the angular distribution of the scattering. From
the relations in Eq.~\ref{eq1a}, one can determine the angular
dependence of the differential cross section in terms of the $j$-scattering
amplitudes. For a circularly
symmetric region where
\begin{equation}
H=H_{0}+V_{T,J,h}
\end{equation}
 for $r<R$, and
\begin{equation}
 H=H_{0}
 \end{equation}
for $r>R$, we have
\begin{eqnarray}\label{diffs}
\sigma_{\tau\tau}(\theta)&=&\frac{2}{\pi
k}\left|\sum^{\infty}_{j\geq\frac{1}{2}}{f_{\tau\tau,j}\cos
j\theta}\right| ^{2}\\ \nonumber
\sigma_{\tau\bar{\tau}}(\theta)&=&\frac{2}{\pi
k}\left|\sum^{\infty}_{j\geq\frac{1}{2}}{f_{\tau\bar{\tau},j}\sin
j\theta}\right| ^{2}\\ \nonumber
\sigma(\theta)&=&\sum_{\tau,\tau'}{\sigma_{\tau\tau'}(\theta)}\; ,
\end{eqnarray}
where $f_{\tau\tau',j}$ are the scattering amplitudes defined in
appendix \ref{appendb}, and $\bar{\tau} = - \tau$.  Notice that the differential cross sections
for  valley-preserving processes, $\sigma_{\tau\tau}(\theta)$,
display a maximum at $\theta=0$, and vanish at $\theta=\pi$ (which
signals the absence of backscattering). The situation is reversed
for the valley-flipping
cross sections, $\sigma_{\tau\bar{\tau}}$, which display a maximum at
$\theta=\pi$, and vanish at $\theta=0$. It is then clear that any
backscattering in the system is caused solely by valley flipping events.

The scattering matrix elements are naturally related to
experimentally measurable quantities via the total, transport and
skew cross sections defined in appendix \ref{appendb}. The
{\em transport} cross section $\sigma_{tr}$ is related to the
transport time
\begin{equation}
\tau_{tr}^{-1}=n_{imp}v_{F}\sigma_{tr}\;,
\end{equation}
and the {\em total} cross section $\sigma_{t}$  to the elastic
scattering time
\begin{equation}
\tau_{e}^{-1}=n_{imp}v_{F}\sigma_{t}\;,
\end{equation}
where $n_{imp}$ is the impurity concentration. The ratio
\begin{equation}
\xi=\frac{\tau_{tr}}{\tau_{e}}=\frac{\sigma_{t}}{\sigma_{tr}}
\end{equation}
reflects the degree
of angular isotropy in the scattering
processes, \cite{helen,scatteringtimes,isotropic}  with $\xi =1$ in the
fully isotropic (massive) limit at low energies.  The
{\em skew} cross section $\sigma_{sk}$, quantifies the
asymmetry of the scattering about the electron's direction of
incidence, and it is related to the appearance of Hall currents,
as characterized through the skew parameter~\cite{ferreira1}
\begin{equation}
\gamma=\frac{\tau_{tr}}{\tau_{sk}}=\frac{\sigma_{sk}}{\sigma_{tr}}\;.
\end{equation}

The form of the differential cross
section in Eq.\ \ref{diffs} results in no skewness, since it is symmetric
about the $x$-axis ($\theta=0$), yielding null skew cross sections (see appendix~\ref{appendb}).

The ratio of scattering times,
\begin{equation}
\xi=\frac{\sum_{\tau\tau'}{\sigma_{t,\tau\tau'}}}{\sum_{tr,\tau\tau'}{\sigma_{\tau\tau'}}}\;,
\end{equation}
can be determined analytically in the low energy
regime $kR\ll1$, where the only contributing channels to the
scattering are the two lowest, $j=\pm\frac{1}{2}$.~\cite{elastic}
Using the relations in Eq.\ \ref{eq:sigmas}, one gets
\begin{equation}\label{eq2}
\xi=\frac{\tau_{tr}}{\tau_{e}}\simeq\frac{2\left(|f_{++,\frac{1}{2}}|^2+|f_{+-,\frac{1}{2}}|^2\right)}{|f_{++,\frac{1}{2}}|^2+3|f_{+-,\frac{1}{2}}|^2}\; .
\end{equation}

Remote charge impurities lead to a local change in the carrier
concentration in their area of influence and a local change of the
Fermi momentum in that region. These impurities can then be
described in the continuum model by the perturbation $vI$,
which respects all the symmetries of the graphene lattice and
belongs to the $A_{1}$ representation of the $C_{6\nu}$ point group.
\cite{disord1,nonabelian} This shift does not lead to valley mixing
and from Eq.\ \ref{eq2} (as $f_{\tau \bar{\tau},\frac{1}{2}} = 0$) one
finds $\xi=2$ for $kR\ll 1$, regardless of the value of $v$, as has been
experimentally shown.\cite{helen,scatteringtimes}
This is a direct consequence and signature of the
anisotropy and predominant forward scattering in graphene.

Some impurities may generate a staggered potential that
differentiates between the $A$ and $B$ sites, as mentioned above,
with a perturbation described by $s\alpha_{3}$ in the continuum model.
This term reduces the point symmetries of the lattice (belongs to the
$B_{2}$ representation of $C_{6\nu}$), and although it does not mix
valleys, it opens a gap. Moreover, a staggered potential
does not commute with $\hat{h}$ or $\mathcal{T}_{1}$,
reducing the set of constraints on the scattering matrix (see
Eq.\ \ref{eqstag}), to those from time reversal and unitarity.
This situation leads to qualitative changes in the isotropy of the scattering
process. The presence of $s\ne 0$ not only affects the ratio of
scattering times $\xi$, but also results in a non-zero skew cross section
(Eq.\ \ref{hallstag}), given that
\begin{equation}
\sigma_{sk,\tau
\tau}=-\sigma_{sk,\bar{\tau}\bar{\tau}}
\end{equation}
and
\begin{equation}
\sigma_{sk,\tau\bar{\tau}}=0\;,
 \end{equation}
and the corresponding appearance of a valley Hall effect.\cite{hallsttagered}
This effect is associated with the appearance of transverse valley currents, which can be characterized by a valley Hall angle
\begin{equation}
\Theta_{VH}=\frac{j_{VH}}{j_{x}}\;.
\end{equation}
At zero temperature the valley Hall angle is equal to the valley skewness, in the absence of side-jump effects (see appendix~\ref{appendb}),
\begin{equation}
\gamma_{V}=\frac{1}{2}(\gamma_{K}-\gamma_{K'})\;.
\end{equation}
This effect is found to be robust to a non-zero $t$ valley-mixing perturbation.
Note that in the case of valley polarized incident currents, $\gamma=\gamma_{V}$, leading to the appearance of Hall voltages due to the
accumulation of charge on the sample edges.

For impurities that accommodate on the center of the hexagon, as is
the case of Ca and Al atoms, \cite{keku}  for a substrate with
a periodicity commensurate with graphene,~\cite{disord3} and for
`Kekul\'e' distortions with zero dimerization
angle,~\cite{disord3} the perturbation terms can be described by $Re
(t \beta e^{i\gamma^{5}\phi}) \simeq t \beta$; this does not break
point symmetries of graphene (belongs to the $A_{1}$
representation of $C_{6\nu}$) but breaks translational invariance,
which leads to vanishing Dirac points and the opening of a
gap.~\cite{disord3} Other hexagon-centered
impurities,\cite{keku,cham1,masses} lead to more general
perturbations with $\phi \ne 0$, as described by $t\beta e^{i\gamma^{5}\phi}$ (in the
$B_{1}$ representation of $C_{6\nu}$), which also open a gap in the spectrum.
The presence of these interactions is reflected in particular through the
deviation of $\xi$ away from $2$ at low energies, as we will describe in detail below.

\begin{figure*}
\includegraphics[scale=0.277]{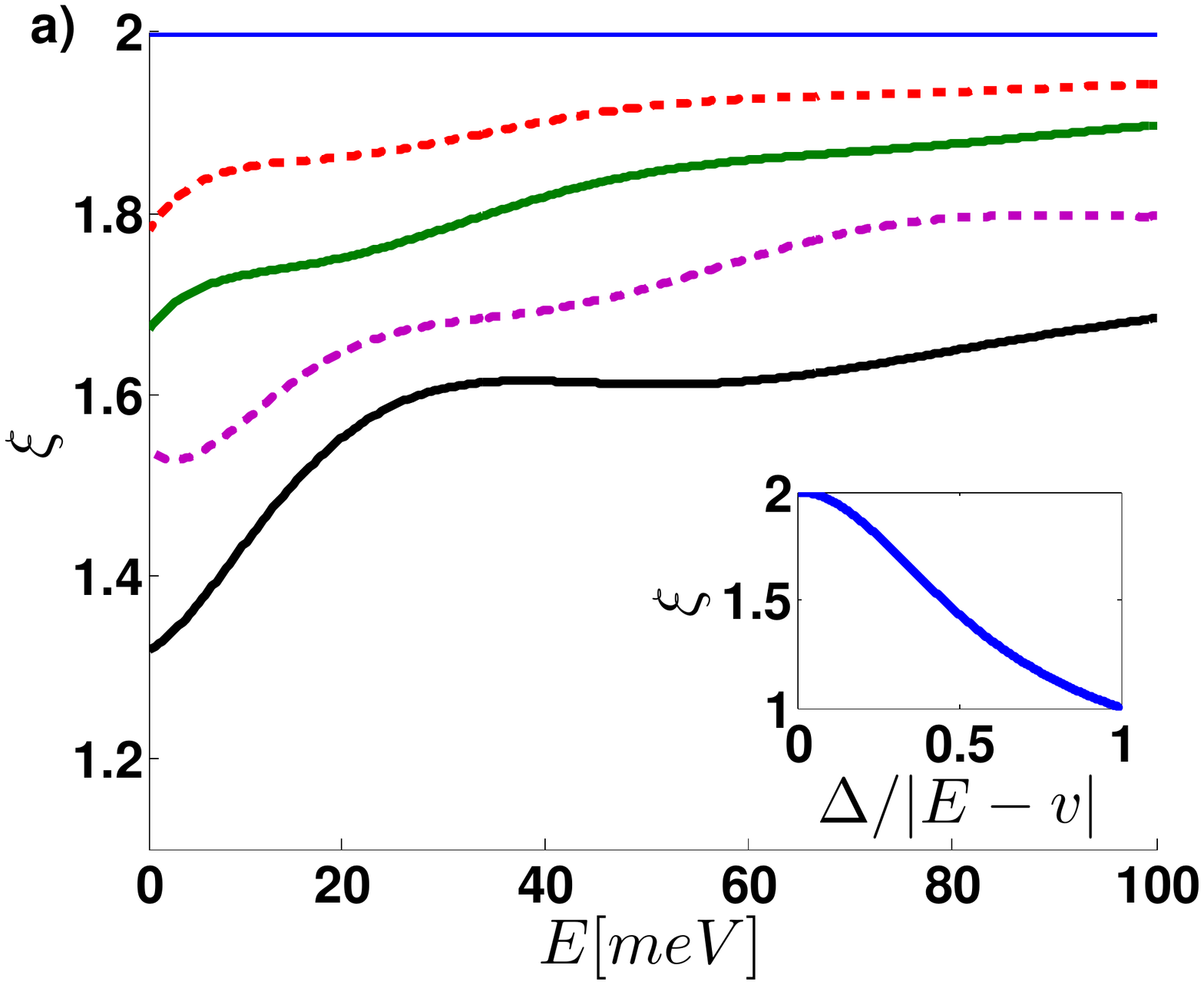}
\includegraphics[scale=0.277]{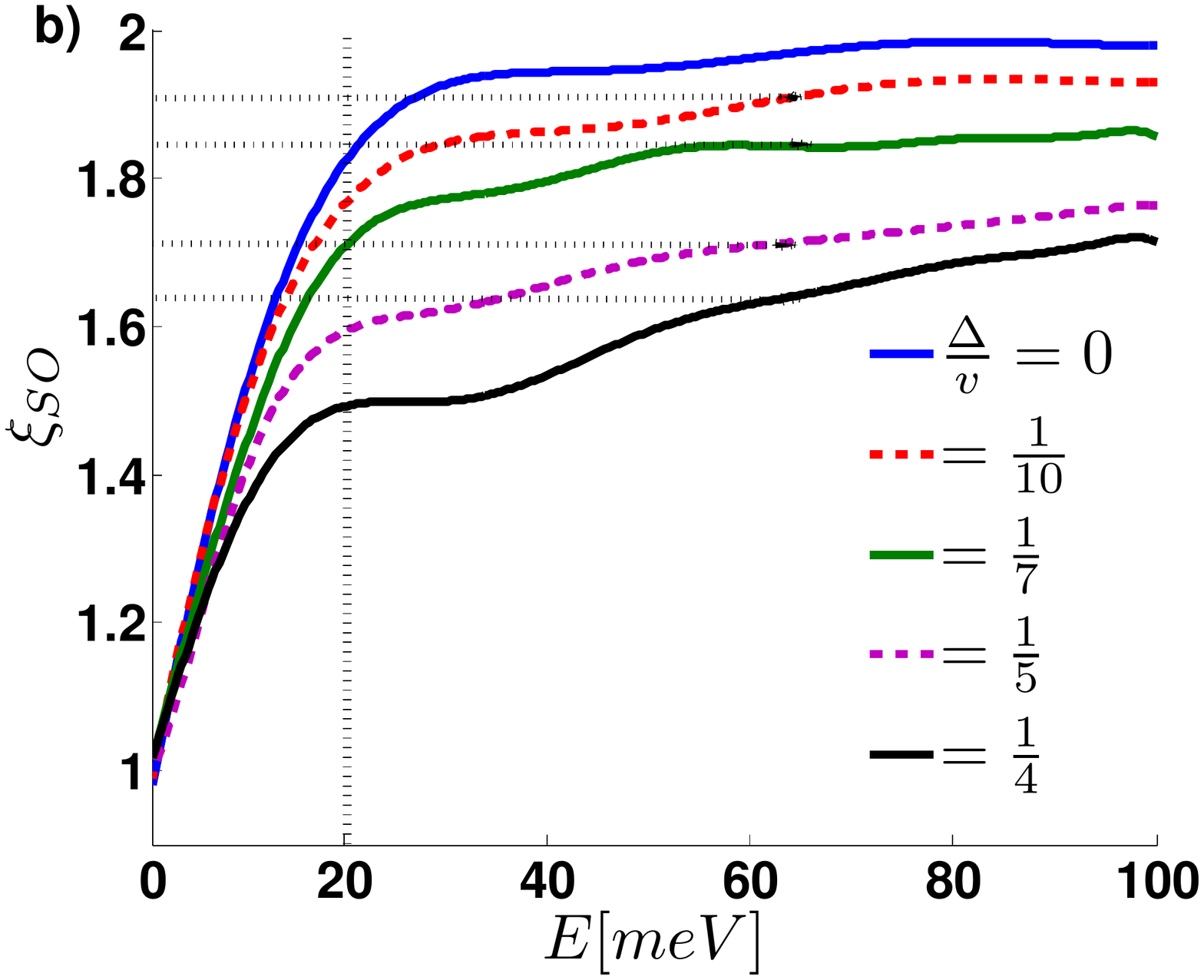}
\includegraphics[scale=0.277]{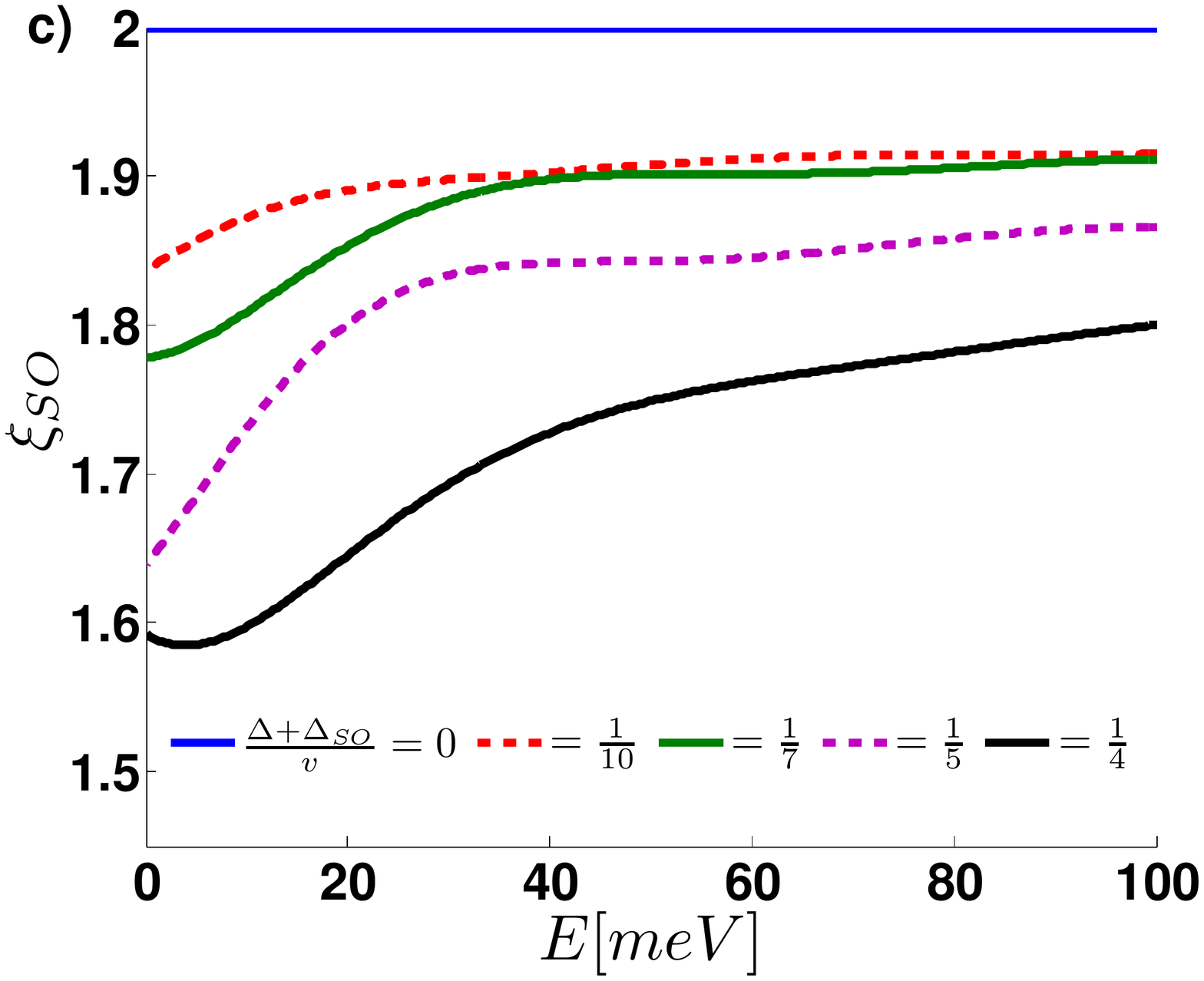}
 \caption{Numerical calculation of the transport to elastic time ratio $\xi$ ($\xi_{SO}$ in the presence of SOIs),
 for a set of $500$ randomly sized impurities, where $(5\le R\le
15)$\AA, and $v=2$eV, for the following systems: a) In the presence
of symmetry breaking effects described by $t$ and $s$,
notice $\xi$ deviates from the
value of $2$ by nearly a constant value at large
energies, and reaches its minimum at/near zero energy.
Inset: Dependence of $\xi$ on $\Delta=\sqrt{t^{2}+s^{2}}$ for
fixed values of $v$ and  $E=60$meV.
b) Effects of $t$ and $s$ on
$\xi_{SO}$ in the presence of a Rashba interaction
$\lambda_{R}=20$meV. Notice the sharp drop of $\xi_{SO}$ at
$E\rightarrow 0$ due to the Rashba interaction. The
deviation from 2 at high energies is due to $t$ and $s$, allowing the
quantification of their amplitudes. Horizontal dashed lines
correspond to the value $\xi_{SO}$ for given $t$ and $s$ in
the inset of a); the vertical dashed line shows the value of
$\lambda_R$; $\xi_{SO}$ drops rapidly for $E < \lambda_R$. c) Effects of $t$ and $s$ on $\xi_{SO}$ in the presence
of intrinsic SOI, $\Delta_{SO}=v/10$. Notice that
$\xi_{SO}$ has a similar behavior as in a),
indicating that the simultaneous extraction of intrinsic SOI and
the symmetry breaking amplitudes $t$ and $s$ from $\xi_{SO}$
is not feasible.} \label{Fig1}
\end{figure*}

In the presence of $t$
and $s$ perturbations, $\xi$ not only depends on these parameters but also
on the value of $v$. Since $\{\alpha_{3},\beta
e^{i\gamma^{5}\phi}\}=0$, and as $\alpha_{i}, \beta$, and $\gamma^{5}$ are
self-inverse (involutory), and obey
$\{\beta,\alpha_{i}\}=\{\alpha_{3},\alpha_{i}\}=0$,
the local spectrum shows a gap of $2\Delta$,
\begin{equation}
k=\sqrt{(E-v)^{2}-\Delta^{2}}\;,
\end{equation}
where
\begin{equation}
\Delta^{2}=s^{2}+t^{2}\;.
\end{equation}

When the potential shift generated by the impurity is such that
$|E-v|\gg |\Delta|$, one recovers the linear (ultra-relativistic
regime), $k\approx |E-v|+\mathcal{O}({\Delta^{2}}/{|E-v|})$,
and the effect of $\Delta$ is minimal on the scattering amplitudes;
this leads to $f_{\tau\bar{\tau},\pm\frac{1}{2}}\approx0$, as well as to
$f_{\tau\tau,\frac{1}{2}}\approx f_{\tau\tau,-\frac{1}{2}}$, and
consequently $\xi\approx 2$, even when $s$ and $t$ may be nonzero.
$\Delta$ should then be comparable to the local kinetic energy,
$\Delta \approx |E-v|$, in order to have a noticeable effect. In this
limit the valley-flipping contributions are non negligible, with
$f_{\tau\tau,\frac{1}{2}}\ne f_{\tau\tau,-\frac{1}{2}}$, and the
ratio $\xi$ exhibits a large drop at low carrier
concentrations, as can be seen in
Fig.\ \ref{Fig1}a.

It should be noted that although the ratio $\xi$ has similar dependence on
energy in the presence of $s$ or $t$, the mechanisms by which
the scattering isotropy is enhanced are different. In the presence
of $t \beta e^{i\gamma^{5}\phi}$, the isotropy is achieved through valley-flip
processes that acquire an amplitude comparable to the valley
preserving processes as $|t|/|E-v|\rightarrow1$, leading to
$\xi\approx 1$ in Eq.~\ref{eq2}. In the case of the staggered perturbation, the
isotropy is enhanced by the decay in amplitude of one
of the two available scattering channels (e.g.
$f_{\tau\tau,\frac{1}{2}}\rightarrow0$ while
$f_{\tau\tau,-\frac{1}{2}}\ne0$  at the K point) as
$|s|/\|E-v|\rightarrow 1$, at each valley
independently.\cite{isotropic}

It is then clear that as graphene is perturbed by adatoms or
deformations, and different symmetry breaking terms appear in the
Hamiltonian, many of the dynamical electronic properties will change
depending on details of the specific terms. In particular, the ratio
$\xi$ will generically deviate from its value of $2$ as the massless
nature of free graphene quasiparticles is affected by these
perturbations. The appearance of backscattering becomes increasingly
important for stronger perturbations and can be detected by
different transport measurements.

\subsection{Spin-orbit interactions} \label{sect:spin}
In the presence of spin-orbit interactions (SOIs) we consider an
extended representation $\psi=(\psi_{A,K,\uparrow},\psi_{B,K,\uparrow},
\psi_{B,K',\uparrow},\psi_{A,K',\uparrow},
\psi_{A,K,\downarrow},\psi_{B,K,\downarrow},
\psi_{B,K',\downarrow},$ $\psi_{A,K',\downarrow})^T$, and define the
corresponding matrices
\begin{eqnarray}
\mathcal{A}_{i}=\alpha_{i}\otimes s_{0}=\left(
                                          \begin{array}{cc}
                                            \alpha_{i} & 0 \\
                                            0 & \alpha_{i} \\
                                          \end{array}
                                        \right) \;, \\
\mathcal{B}=\beta\otimes s_{0}=\left(
                                 \begin{array}{cc}
                                   \beta & 0 \\
                                   0 & \beta \\
                                 \end{array}
                               \right)\;, \\
\Gamma^{5}=\gamma^{5}\otimes s_{0}=\left(
                                     \begin{array}{cc}
                                       \gamma^{5} & 0 \\
                                       0 & \gamma^{5} \\
                                     \end{array}
                                   \right)\;,
\end{eqnarray}
which operate on the pseudospin and valley degrees of
freedom, but leave the spin unchanged, as $s_{0}$ is the identity in
spin space. These operators satisfy
\begin{eqnarray}
\{\mathcal{A}_{i},\mathcal{A}_{j}\}&=&2i\delta_{ij}\;,\\
\{\mathcal{B},\mathcal{A}_{i}\}&=&0 \;, \\
\left[\mathcal{A}_{i},\Gamma^{5}\right]&=&0 \;,
\end{eqnarray}
 and
\begin{equation}
\{\mathcal{B},\Gamma^{5}\}=0\;.
\end{equation}
We also define
\begin{equation}
S_{i}=s_{i}\otimes\sigma_{0}\otimes \tau_{0}\;,
\end{equation}
which act on the spin degree of
freedom, with
\begin{equation}
\{S_{i},S_{j}\}=2\delta_{ij}\;,
\end{equation}
and commute with
$\mathcal{A}_{i}$, $\mathcal{B}$ and $\Gamma^{5}$. In this
representation, the time reversal,\cite{newsym} helicity operator, and total
angular momentum are defined as
\begin{eqnarray}
\mathcal{T}&=&i\mathcal{B}\Gamma^{5}\mathcal{A}_{1}S_{2}\mathcal{C}\;,\\
\hat{h}&=&\Gamma^{5}\vec{\mathcal{A}}\cdot \vec{k}/k\;,\\
J_{z}&=&-i\hbar\partial_{\theta}+\Gamma^{5}\mathcal{A}_{3}/2+S_{3}/2\;,
\end{eqnarray}
with $J_{z}$ having integer eigenvalues, $\mathcal{T}^2=-1$, and $\{\mathcal{T},J_z \} =0$. Similarly, one can define'
\begin{equation}
\mathcal{T}_{1}=e^{-i\Gamma^{5}\phi}\Gamma^{5}\mathcal{A}_{2}
S_{2}\mathcal{C}\;,
\end{equation}
with $\mathcal{T}_1^2 = 1$, and satisfying
\begin{equation}
\{\mathcal{T}_1 , J_z\} = 0\;.
\end{equation}
\subsubsection{Symmetry considerations and constraints on spin transport}
In this representation, the set of spin dependent
interactions that conserve total angular momentum is found to be
\begin{multline}
V^{SO}_{T,J}=\{\lambda_{R}(\mathcal{A}_{1}S_{2}-\mathcal{A}_{2}S_{1}),\lambda(\mathcal{A}_{2}S_{2}+\mathcal{A}_{1}S_{1}),\\ \Delta_{SO}\Gamma^{5}\mathcal{A}_{3}S_{3},\Delta_{S}\Gamma^{5}S_{3}\}\;,
\end{multline}
all with real coefficients.  These interactions commute with both $\mathcal{T}$ and $\mathcal{T}_1$.
For
\begin{equation}
H=H_{0}+V_{\mathcal{T}_{1}}\;,
\end{equation}
where
\begin{equation}
V_{\mathcal{T}_{1}}=V^{SO}_{T,J} \cup \{v I,
t \mathcal{B}e^{i\Gamma^{5}\phi}\}\;,
\end{equation}
one finds
\begin{equation}
\left[\mathcal{T}_{1},H]=[\mathcal{T},H\right]=0,
\end{equation}
and
utilizing the unitarity of the $S$ matrix one obtains the condition
\begin{equation}
\Gamma^{5}\mathcal{T}_{1}S_{j}\mathcal{T}_{1}^{-1}\Gamma^{5}=S^{\dag}_{-j}
\end{equation}
(see Eq.\ \ref{eq:C5}).

It should be noted that the SO terms $\lambda_{R}$ and $\lambda$ do
{\em not} lead to a gap opening; instead they break the spin degeneracy by
doubling the bands for a fixed energy, such that
\begin{equation}
k_{\pm}=\sqrt{(E-v)^{2}\pm2\Upsilon(E-v)}\;,
\end{equation}
where
\begin{equation}
\Upsilon=\sqrt{\lambda^{2}_{R}+\lambda^{2}}\;.
 \end{equation}
 These interactions
split the low energy resonant peaks corresponding to $j=0$ and
$j=\pm1$ channels whenever $\Upsilon>2E$,~\cite{isotropic} and
produce a noticeable drop at low carrier concentrations
($E<\Upsilon$) in the ratio of cross sections,
\begin{equation}
\xi_{SO}=\frac{\sum_{\tau\tau'ss'}{t,\sigma_{\tau\tau'ss'}}}{\sum_{\tau\tau'ss'}\sigma_{tr,\tau\tau'ss'}}\;,
\end{equation}
as shown in Fig.~\ref{Fig1}b.

The spin independent terms lead also to a
variation of $\xi_{SO}$ at low energies, although with a different
dependence on the carrier concentration, as shown in
Fig.~\ref{Fig1}a. Therefore, in the presence of $\lambda_{R}$,
$\lambda$, $t$ and $s$ terms one can determine the value of
$\upsilon$ and the ratio $\Delta/v$ from the behavior of $\xi_{SO}$ with Fermi energy,
as illustrated in Fig.~\ref{Fig1}b.

The intrinsic SOI
($\Delta_{SO}$) does not commute with
$\mathcal{T}_{1}$ and does not cause spin flips, but opens a gap, and leads
to a reduction of $\xi_{SO}$, such that $\xi_{SO}\rightarrow1$ as
$\Delta_{SO}/|E-v|\rightarrow 1$. However, the gap generated by
$\Delta_{SO}$ is incompatible with the gaps generated by
$t$ and $s$, since these interactions do not anticommute with the intrinsic
SOI.~\cite{masses}  In this case, the effect of the intrinsic SOI can
not be distinguished from symmetry breaking effects in samples
through measurements of the ratio $\xi_{SO}$.  In the presence of $\Delta_{SO}$ and with no symmetry breaking terms ($t=s=0$),
the dependence of $\xi_{SO}(E)$ on the different values of $\Delta_{SO}$ are qualitatively similar to those in Fig.~\ref{Fig1}a.
It is the case that the simultaneous presence of $\Delta_{SO}$,  $t$ and $s$ will lead an overestimation of $\Delta_{SO}$ from $\xi_{SO}$
measurements, due to the similar deviation caused by $t$ and $s$, as shown in Fig.\ \ref{Fig1}c.

\subsection{Skew scattering and spin Hall effect } \label{sect:skew}
An alternative method to measure the strength of SOIs in graphene was
introduced recently by Ferreira {\em et al}.\cite{ferreira1} The
method exploits the skew character of scattering in the presence of
locally enhanced SOIs. Here, we study effects produced by dimerization perturbations such
as $t\mathcal{B}e^{i\Gamma^{5}\phi}$, and staggered sublattice symmetry
breaking, and how they influence  skew scattering and the possible determination of the SOI parameters from transport experiments.

In the absence of valley mixing effects or sublattice symmetry
breaking, the $V^{SO}_{T,J}$ perturbations commute with both $\mathcal{T}$ and
$\mathcal{T}_{1}$, as mentioned above. The symmetries of the scattering matrix
result in $\sigma_{sk, \tau\tau \uparrow\uparrow}=-\sigma_{sk,
\tau\tau\downarrow\downarrow}$ and $\sigma_{sk, \tau\tau
\uparrow\downarrow}=-\sigma_{sk, \tau\tau\downarrow\uparrow}=0$ (see appendix~\ref{append3}).
This indicates that carriers with opposite spin are scattered
in opposite directions, which would lead to their
accumulation on the sample edges and the appearance of
a spin Hall effect (SHE) signal. \cite{ferreira1}

One way to quantify this effect is through the ratio of the skew
to transport cross section for each spin
projection, $\gamma_{\uparrow}$ and $\gamma_{\downarrow}$, and define
the spin transport skewness as
\begin{equation}
\gamma_{S}=\frac{1}{2}(\gamma_{\uparrow}-\gamma_{\downarrow})
\end{equation}
where
\begin{equation}
\gamma_{s}=\frac{\sum_{\tau\tau's'}{\sigma_{sk,\tau\tau',s's}}}{\sum_{\tau\tau's'}{\sigma_{tr,\tau\tau's's}}} \, ,
\end{equation}
(see Eq.~\ref{skewparmater} in appendix), and $\gamma_{S}$ considers valley and
spin preserving events, as well as valley and spin reversals.
The spin skewness $\gamma_{S}$ considers then all processes resulting in spin asymmetries,
it is enhanced through resonant scattering, and fully characterizes the
spin Hall angle at zero temperature,
\begin{equation}
\theta_{SH}=\left. \frac{j_{SH}}{j_{x}} \right |_{T=0}=\gamma_{S} \, ,
\end{equation}
in the absence of side-jump effects.~\cite{ferreira1}
In the absence of valley mixing or sublattice symmetry
breaking, the skew parameters for each spin are equal in
magnitude and opposite in sign, leading to
$\gamma_{S}=\gamma_{\uparrow}$. This effect was recently observed in
graphene samples in which the SO was enhanced through the deposition
of large copper or gold nanoparticles,\cite{nanoparticles} and in
weakly hydrogenated samples, and found to yield rather strong SHE
signals.~\cite{colossal}

\begin{figure}
\includegraphics[scale=0.2]{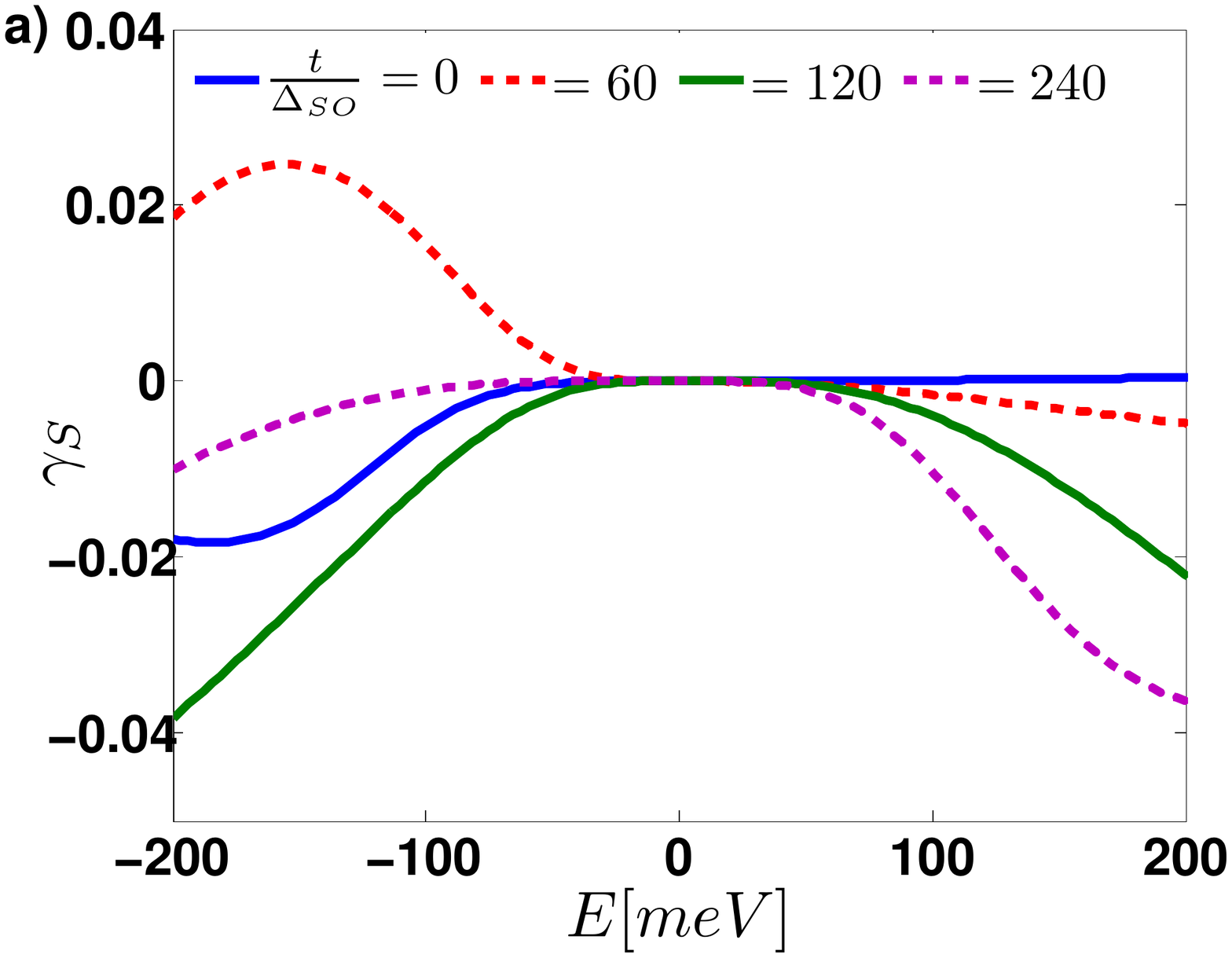}
\includegraphics[scale=0.2]{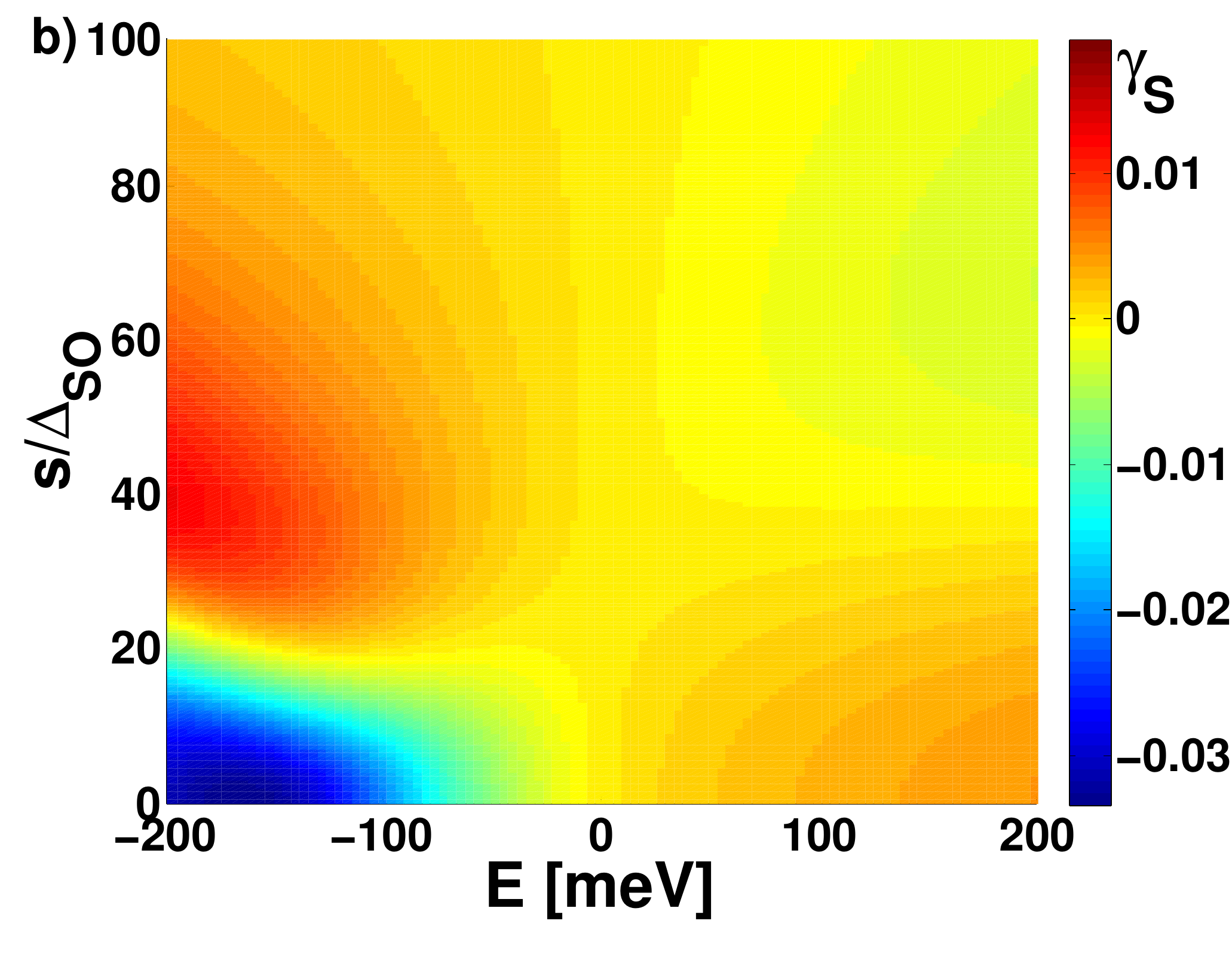}
\includegraphics[scale=0.2]{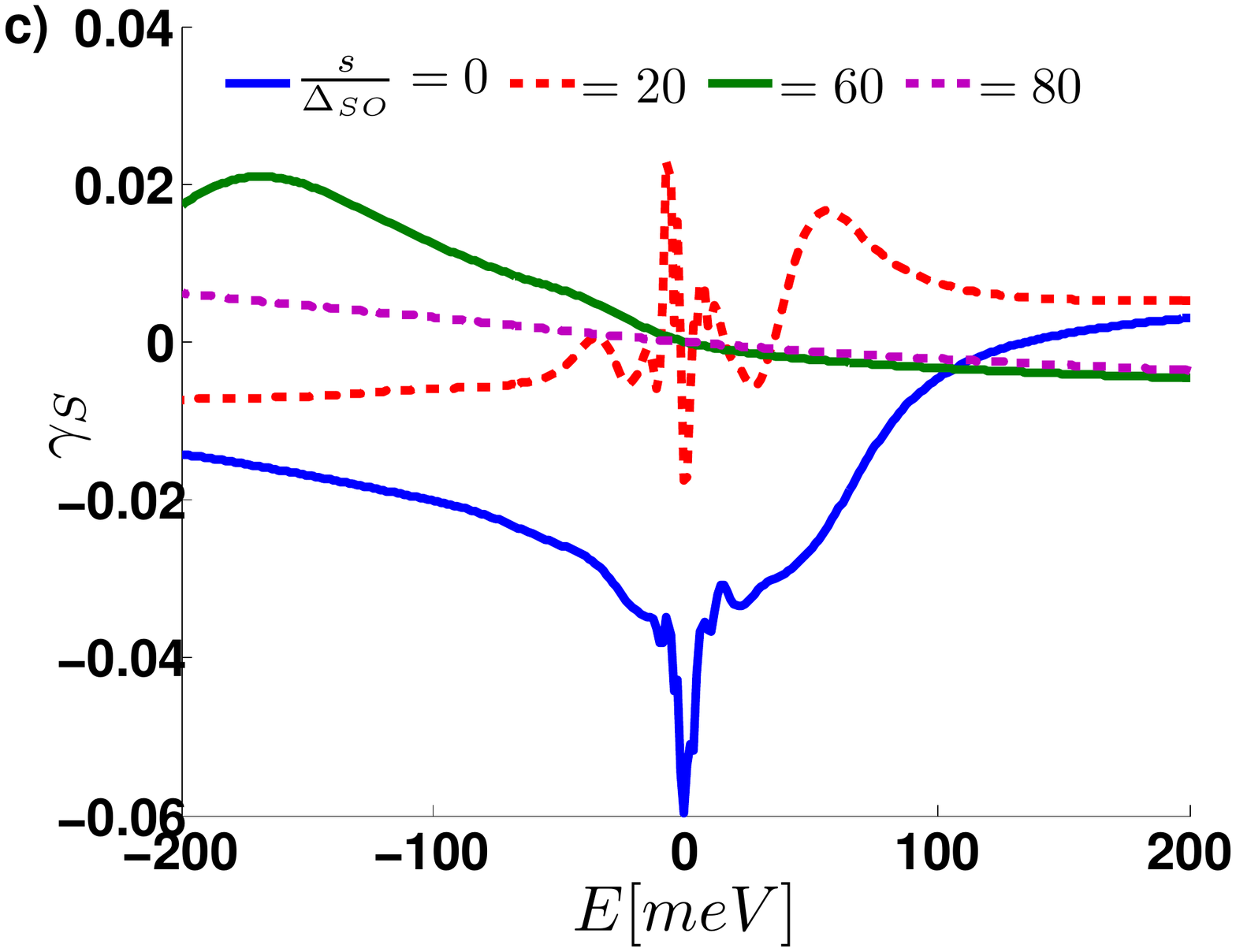}
\includegraphics[scale=0.2]{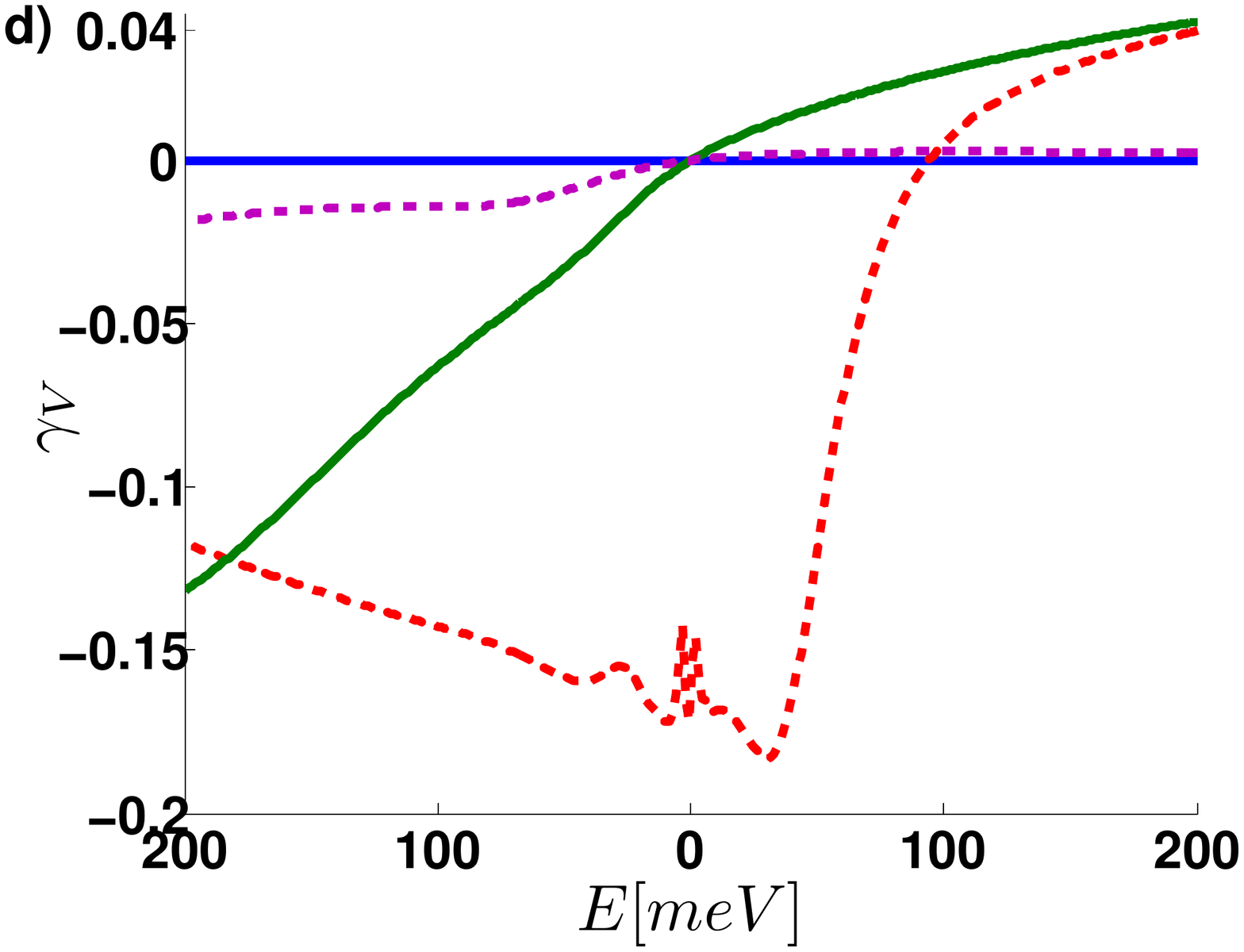}
 \caption{a) Spin skew parameter $\gamma_{S}$ as function of energy for different values of $t$
 in a system with $\Delta_{SO}=25$meV, $v=2$eV and
 $R=6$\AA. Even for large values of $t$ the skew scattering persists, as
expected from symmetry arguments. b) Map of spin skew parameter $\gamma_{S}$
vs symmetry breaking staggered interaction $s$ and energy
(carrier concentration) $E$, for
 $\Delta_{SO}$, $v$ and $R$ as in a). Notice that for large values of $s$ the skew parameter $\gamma \rightarrow
 0$, showing that skew scattering is not robust to this kind of
 perturbation.  c) Spin skew parameter dependence
 on $s$ for a set of $2000$ randomly sized impurities $(3\le R \le 9)$\AA, for values of $v$ and $\Delta_{SO}$ as in (a).
 This shows the robustness of skew scattering to
 random impurity size, but not under $s$ perturbations. d) Valley skew parameter dependence
 on $s$ for the same set of random impurities and $s$ values in (c). Note that in systems with $s>\Delta_{SO}$, it is generally the case 
 that $|\gamma_{V}|>|\gamma_{S}|$ (notice different vertical scales).
 Both $\gamma_{S}$ and $\gamma_{V}$ vanish away from the resonant regime, as it is the case for $s/\Delta_{SO}=80$.} \label{Fig2}
\end{figure}

In the presence of valley mixing,
$t\mathcal{B}e^{i\Gamma^{5}\phi}$, in addition to SOI terms, both $\mathcal{T}$ and $\mathcal{T}_{1}$
commute with the Hamiltonian, resulting in the following relations,
$\sigma_{sk,\tau\tau,ss}=-\sigma_{sk,\tau\tau,\bar{s}\bar{s}}=-\sigma_{sk,\bar{\tau}\bar{\tau},\bar{s}\bar{s}}=\sigma_{sk,\bar{\tau}\bar{\tau},ss}$
and $\sigma_{sk,\tau\tau',s\bar{s}}=\sigma_{sk,\tau\tau',\bar{s}s}=0$,
as well as a nonzero value of the skew parameter $\gamma$;  its dependence
on $t$ as a function of carrier concentration is illustrated in Fig.~\ref{Fig2}a.

However, in the presence of a staggered perturbation
$s\mathcal{A}_{3}$, the Hamiltonian does not commute with the
operator $\mathcal{T}_{1}$.  This reduces the set of conditions
imposed on the skew cross section elements, so that the relation between
$\sigma_{sk,\tau\tau,ss}$ and $\sigma_{sk,\tau\tau,\bar{s}\bar{s}}$
depends on the staggered parameter $s$ (see Eqs.\ \ref{skewspinsttagered}
and \ref{skewparmater}).
A nonzero $s$ leads to a
reduction of $\gamma_{S}$ and the corresponding transverse spin currents.
Figure \ref{Fig2}b shows the energy dependence of $\gamma_{S}$ for
different values of the staggered perturbation $s$ in the system of Fig.\ \ref{Fig2}a.
For increasing values of $s$ the spin skew
parameter $\gamma_{S}$ decreases and eventually vanishes for large $s$
values. In Fig.\ \ref{Fig2}c we show the effect of
$s$ on a sample of randomly sized impurities. We
notice that the SHE is robust to impurity size disorder, but it is {\em not} to staggered potential
perturbations; $\gamma_{S}$ decreases with increasing values of
$s$, eventually vanishing for all Fermi energies.  It is also interesting that the presence of $s$
leads a non-zero valley skewness, $\gamma_{V}$, which would then result in a valley Hall effect, accumulating carriers from different valleys
to different edges of the sample.  Figure \ref{Fig2}d shows the dependence of $\gamma_V$ on energy and $s$ strength,  for a
random impurity configuration, reflecting a relatively large value of $\gamma_V$ for sizable $s$, while $\gamma_S$ is much smaller.
At high values of $s$, however, the resonance conditions are offset and both $\gamma_V$ and $\gamma_S$ are small in the
energy range shown.

\begin{figure}
  \centering
  \includegraphics[scale=0.4]{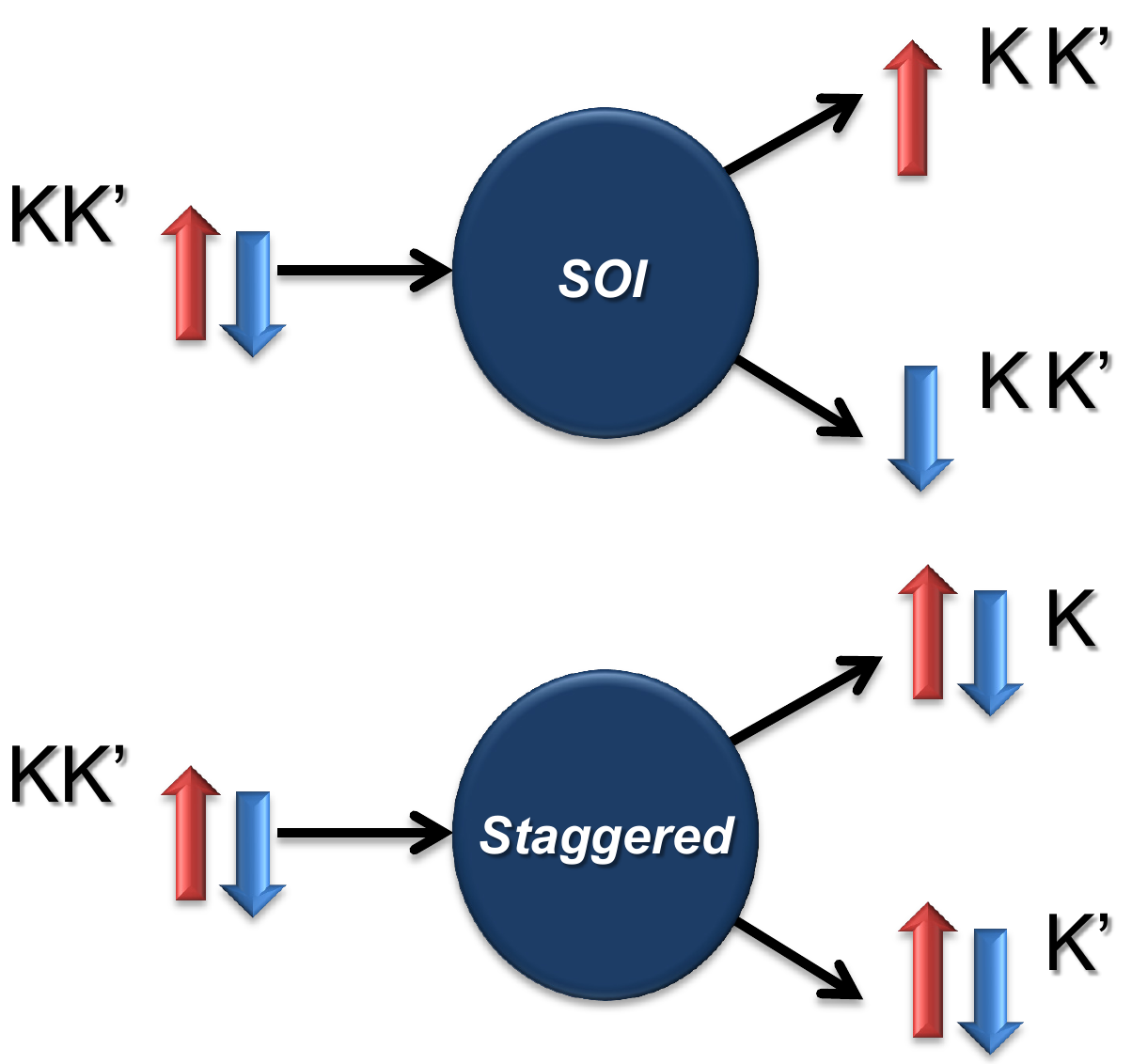}\\
  \caption{Schematic representation of spin and valley skew scattering processes due to different perturbations.
SOI perturbations act as a spin splitter while leaving the valleys unaffected. The skew character of the scattering remains unaffected in the presence of
$V_{\mathcal{T}_{1}}$. In contrast, the presence perturbations that belong to $V_{\mathcal{T},J}$ and do not commute with $\mathcal{T}_{1}$,
such as the staggered potential $s$, result in valley skewness. In this case the perturbation acts as a valley splitter, while leaving the spin unaffected.
In the presence of a staggered potential and SOIs, spin skewness will be reduced, depending on the strength of the staggered effects. }\label{scem}
\end{figure}

We emphasize that SHE appears and it is enhanced close to
the resonant scattering regime, \cite{ferreira1} due to the
presence of centrosymmetric impurities that enhance
SOIs (and produce also a local potential shift).
This effect is robust to impurity size and potential disorder,
\cite{ferreira1} as well as to perturbations  that lead to valley mixing
in the form $t\mathcal{B}e^{i\Gamma^{5}\phi}$. Notice also that close to resonance, both the spin and valley Hall effects are enhanced,
but become negligible far away from resonance,
as is the case for $s/\Delta_{SO}=80$ in Fig.\ \ref{Fig2}c and d.
The competition between these two effects can be schematically illustrated as in Fig.\ \ref{scem}:
the SHE generated by skew scattering is not robust to a staggered potential, as this leads to a non-zero
valley skewness.  The latter reduces spin accumulation on the sample edges, while enhancing the valley contrast.

\section{Discussion}  \label{sect:discussion}

In light of our results and examination of symmetry breaking effects, let us consider recent experimental results on decorated graphene.

In the experiments by Monteverde {\em et al}., \cite{helen}  graphene was deposited on SiO$_{2}$, and a ratio of transport to elastic times $\xi\approx 1.8$ was observed.
The lowest carrier concentration for which $\xi$ is reported corresponds to $E\approx 100$ meV (see their Fig.\ 3).  As $\xi$ did
not display a sharp drop below the value of $2$ until the experimental minimum $E$ considered, we can safely consider that the value of any Rashba SOI is lower than
$\approx 200$ meV, if present. \cite{isotropic} [It is interesting to notice that sample A in their Fig.\ 3 exhibits a drop in the value of $\xi$ for both polarities near the minimum
carrier density, suggesting the presence of Rashba SOI in that sample.]  We can also neglect the intrinsic SOI in pristine graphene, \cite{intrinsic} and it being unlikely enhanced
by the SiO$_{2}$ substrate. \cite{indium} However, the effects of substrate and impurities are more likely to be symmetry breaking, with a staggered $s$,
and/or hopping modulation $t$, character. Assuming this to be the case, and comparing with the results of Fig.\ \ref{Fig1}b, we estimate the value of
$\Delta/v$ (with $\Delta=\sqrt{t^{2}+s^{2}}$) to be $\Delta/v\approx 1/7$, or $\Delta \approx 300$meV -- not an unreasonable estimate of the staggered perturbation strength.

In the experiments by Marchenko {\em et al}.,\cite{Nii}
Rashba interaction was enhanced through the intercalation of gold
atoms between graphene and the Ni substrate. The
value  of the Rashba interaction was found to depend on the detailed position of the gold atoms with
respect to the graphene sample.  For samples exhibiting a Rashba
interaction of $\sim70$meV, gold atoms  were located in the center
of the hexagon; for $\sim60$meV, the gold atoms were in the center of
non consecutive hexagons; and for $\sim7$meV, the gold atoms were on
top of carbon locations. Each of these arrangement can be seen to
generate different symmetry breaking terms, in addition to the
enhanced SOI.\@  In the case of hexagon-centered gold atoms one expects
the perturbation to be dominated by $t$ and $v$, as
these terms possess the symmetries of this specific arrangement. In the
cases of $60$ and $7$meV, one can expect the symmetry breaking to be described
by $\{t, s, v\}$ and $\{s,v\}$, respectively. In
those cases, the detection and quantification of the Rashba SOI
based on transport and elastic times not only allows one to quantify
the strength of the Rashba interaction but also to obtain an estimate
of the additional parameters of the Hamiltonian (see Fig.~\ref{Fig1}b).

For example, in the case of a Rashba interaction
$\lambda_{R}=20$meV, we notice a sharp drop in the ratio
$\xi_{SO}$ for a Fermi energy around this value (indicated by a dashed vertical
line) in Fig.~\ref{Fig1}b. For values of energy larger than $20$meV, the ratio displays
a nearly constant value, slightly shifted down from
the value of $2$ (as indicated by horizontal dashed lines in
Fig.~\ref{Fig1}b), with an  asymptotic value dependent on the value of $\Delta/v$.

The decoration of graphene with heavy adatoms positioned at
the center of hexagons leads to the enhancement of intrinsic
SOI.\cite{indium} The detection and quantification of this
interaction through transport measurements becomes masked by
symmetry breaking effects, as shown in Fig.\ \ref{Fig1}c. This would
lead to overestimation of $\Delta_{SO}$ from measurements of $\xi_{SO}$ with
varying Fermi energy. This
suggests that different measurements, such as SHE
should be used in conjunction in order to fully characterize the role of
the adatom perturbation.

Skew scattering and the resulting SHE represents an alternative
method of quantification of SOIs. The
physisorption of nanoparticles on the surface of
graphene  may be expected to produce a potential
shift $v$ and enhanced SOIs  with no additional symmetry breaking
effects, given their large size when compared to the graphene lattice constant,
leading to more accurate quantification of the
SOIs.~\cite{nanoparticles}
Intrinsic SOI enhanced by heavy
adatoms,~\cite{indium} may be characterized through this method.
One can expect staggered effects to be weak in these samples, $s \ll \Delta_{SO}$,
so that skew scattering would be an important contribution to the SHE.

Our results are also in general agreement with the theoretical analysis by Pachoud {\em et al.}, \cite{ferreira2}
who examine the different low-energy perturbations arising from different adatom locations on the graphene lattice.
They notice that the Hall effects produced from hexagon centered or on-top adatoms are of different character. For the hexagon centered position
they expect a pure spin Hall effect, in agreement with the symmetry constrains presented in our work.  Moreover, for on-top positions it was suggested that if the adatom
number of on-$A$ and on-$B$ types is different, a Hall effect may be observed.  This is in agreement with our results for the staggered potential perturbation 
and the corresponding appearance of a valley Hall effect, which could eventually yield a charge Hall effect for valley polarized currents.

Finally, in hydrogenated samples, where the Rashba interaction is
enhanced due to $sp^{3}$ deformations,~\cite{colossal,hydrogen}
the non-local resistance was seen to be independent of carrier
concentration. This suggested that skew scattering was negligible,
and the appearance of SHE was attributed to the side-jump
mechanism.~\cite{colossal} The weak contribution of the skew
scattering in this kind of sample, however, may also be due
to the presence of strong symmetry breaking effects that reduce the
conditions for the appearance of SHE, such as a strong staggered
perturbation, $s\gg \lambda_{R}$. This regime would result in a
negligible spin skew parameter, as shown in Fig.~\ref{Fig2}c, due to the dominance of the valley skewness.
The persistence of SHE would further suggest that side jump effects are robust to this kind of perturbations, and/or that the phenomenon observed
should consider the role of valley skewness in the non local resistance. It would be interesting to carry out a detailed study of this issue in the future.

\section{Conclusions} \label{sect:conclude}
We have studied how symmetry breaking effects lead to the appearance of backscattering
in graphene, which is reflected among other things on the deviation of the transport to
elastic time ratio $\xi$ from its ideal value of
$2$ at low carrier concentrations.
We have also considered that perturbations introduced by adatoms in graphene lead to the
enhancement of Rashba-like interactions, in addition to various symmetry
breaking effects.  In many situations, we find it is
possible to detect and separately quantify the SOI and the symmetry
breaking amplitudes through measurements of the ratio of transport and energy
relaxation times, $\xi$, as function of carrier concentration. This separability of contributions becomes a challenging
task if the SOI enhancement has an intrinsic character, however. In that case, the dependence of $\xi$
on carrier concentration for SOI and symmetry breaking interactions is quite similar, obstructing the extraction of each effect
independently. Finally, we discussed that nonzero spin and/or valley skew parameters,
$\gamma_{S}$ and $\gamma_{V}$, are expected for certain adatoms. This allows
the detection of both Rashba and intrinsic SOIs in the
presence of valley mixing effects, even when the transport times ratio may not give definitive answers.
For adatoms that enhance SOIs and break
sublattice symmetry, we find a reduction of $\gamma_{S}$, which vanishes in systems where the staggered perturbation is larger that
the SOI amplitude. As a consequence, the SOI strength would be masked by sublattice symmetry breaking, and the appearance of a valley Hall effect.

\begin{acknowledgments}
We thank N. Sandler and C. Lewenkopf for helpful discussions.
MMA is grateful to the 2014 Les Houches Summer School {\em Topological Aspects of Condensed Matter Physics},
for the welcoming and enlightening environment. This work was supported in part by
NSF CIAM/MWN grant DMR-1108285.
\end{acknowledgments}

\appendix
\section{Spinless scattering matrix}\label{appenda}
In the presence of the $V_{T,J,h}$ interactions the operator
$\mathcal{T}_{1}=ie^{-i\gamma^{5}\phi}\gamma^{5}\alpha_{2}\mathcal{C}$
and time reversal operator
$\mathcal{T}=\beta\gamma^{5}\alpha_{1}\mathcal{C}$, where
$\mathcal{C}$ is the complex conjugation operator, commute with the
total Hamiltonian. In addition, $\mathcal{T}^{2}_{1}=-1$,
$\mathcal{T}^{2}=1$,
$\{\mathcal{T}_{1},J_{z}\}=\{\mathcal{T},J_{z}\}=0$,
$[\mathcal{T}_1,\hat{h}]=0$, $\{\mathcal{T},\hat{h}\}=0$, which
allows one to choose the states $j$ and $-j$, such that
$\mathcal{T}_{1}\psi_{\pm,j}=\pm(-1)^{j}\psi_{\pm,-j}$ and
$\mathcal{T}\psi_{\pm,j}=\mp(-1)^{j+\frac{1}{2}}\psi_{\mp,-j}$. Note
that in cylindrical coordinates one can explicitly write eigenstates
that explicitly satisfy these relations, such as $\psi_j =
(J_{j-1/2}, iJ_{j+1/2} \, e^{i\theta} , J_{j-1/2} , -iJ_{j+1/2}\,
e^{i\theta})^T e^{i(j-\frac{1}{2})\theta}$, for $E>0$, and where
$J_l (kr)$ are Bessel functions.
 \cite{isotropic}

In order to study the scattering problem, the incoming and outgoing states are
eigenstates of the free Hamiltonian, $H_{0}\psi=E\psi$. These eigenstates can be represented
with the help of the (helicity) projection operators $P_{\pm}=(1\pm\gamma^{5})/2$,
so that a general eigenstate of the Hamiltonian can be written as
\begin{equation}
\psi=P_{+}\psi+P_{-}\psi
\end{equation}
with $[\hat{h},P_\pm]=0$ and
$\hat{h}(P_{\pm}\psi)=\pm P_\pm \psi$. In addition, for a partial wave component we have
\begin{eqnarray} \label{eq:T1psij}
\mathcal{T}_{1}\psi_{j}&=&\mathcal{T}_{1} P_{+}\psi_{j} + \mathcal{T}_{1} P_{-}\psi_{j}\; \nonumber \\
 &=&(-1)^{j}(P_{+}-P_{-})\psi_{-j} \;  \nonumber \\
 &=&(-1)^{j}\gamma^{5}\psi_{-j}\; .
\end{eqnarray}
Similarly, for the operator $\mathcal{T}$, one gets
\begin{equation}
\mathcal{T}\psi_{j}=(-1)^{j+\frac{1}{2}}\gamma^{5}\psi_{-j}\; .
\end{equation}

The scattering of a single partial wave
$j$ can be written in terms of incoming $\psi^{in}_{j}$ and outgoing
$\psi^{out}_{j}$ radial eigenstates of the free Hamiltonian,
\begin{equation}\label{sj}
\psi_j = \psi^{in}_{j}+S_{j}\psi^{out}_{j}\; ,
\end{equation}
where,
\begin{eqnarray}
S_{j}&=&\left(
                      \begin{array}{cc}
                        \hat{S}_{++,j} & \hat{S}_{+-,j}\\
                        \hat{S}_{-+,j} & \hat{S}_{--,j} \\
                      \end{array}
                    \right)\; ,  \\
                     \nonumber
\hat{S}_{j}&=&S_{\tau\tau',j}\sigma_{0}   \; .
\end{eqnarray}

Applying the operator $\mathcal{T}_{1}$ to the partial  wave
component $\psi_{j}$ and using \ref{eq:T1psij} we have

\begin{eqnarray}\label{j}
&\mathcal{T}_{1}\psi_{j}=\mathcal{T}_{1}\psi^{in}_{j}+\mathcal{T}_{1}S_{j}\mathcal{T}^{-1}_{1}\mathcal{T}_{1}\psi^{out}_{j}\\
\nonumber
\end{eqnarray}
or
\begin{equation}
\psi_{-j}=\psi^{out}_{-j}+\gamma^{5}\mathcal{T}_{1}S_{j}\mathcal{T}^{-1}_{1}\gamma^{5}\psi^{in}_{-j}\; .
\end{equation}
Since the $S$ matrix takes a block diagonal form, unitarity  $SS^{\dag}=1$,
requires $S_{j}S^{\dag}_{j}=1$. Comparison with the $-j$ wave component,
\begin{equation}\label{minusj}
\psi_{-j}= \psi^{in}_{-j}+S_{-j}\psi^{out}_{-j} \, ,
\end{equation}
results in the following condition for $S_j$,
\begin{equation}
\gamma^{5}\mathcal{T}_{1}S_{j}\mathcal{T}^{-1}_{1}\gamma^{5}=S^{\dag}_{-j} \, ,
\end{equation}
or equivalently,
\begin{equation}
\left(
  \begin{array}{cc}
    \hat{S}_{++,j} & -\hat{S}_{+-,j}e^{2i\phi} \\
    -\hat{S}_{-+,j}e^{-2i\phi} & \hat{S}_{--,j} \\
  \end{array}
\right)=\left(
          \begin{array}{cc}
            \hat{S}_{++,-j} & \hat{S}_{-+,-j} \\
            \hat{S}_{+-,-j} & \hat{S}_{--,-j} \\
          \end{array}
        \right)\; .
\end{equation}

Similarly, symmetries under $\mathcal{T}$ result in
\begin{equation}
\gamma^{5}\mathcal{T}S_{j}\mathcal{T}^{-1}\gamma^{5}=S^{\dag}_{-j}\, ,
\end{equation}
or
\begin{equation} \label{eq:SwithT}
 \left(
  \begin{array}{cc}
    \hat{S}_{--,j} & -\hat{S}_{-+,j} \\
    -\hat{S}_{+-,j}  &  \hat{S}_{++,j} \\
  \end{array}
\right)=\left(
          \begin{array}{cc}
            S_{++,-j} & S_{-+,-j} \\
            S_{+-,-j}& S_{--,-j} \\
          \end{array}
        \right) \, ,
\end{equation}
with the resulting relations
\begin{eqnarray}\label{eq1a}
&S_{\tau\tau,j}=S_{\tau\tau,-j}=S_{\bar{\tau}\bar{\tau},-j}=S_{\bar{\tau}\bar{\tau},j}\; , \\ \nonumber
&S_{\tau\bar{\tau},j}=-S_{\bar{\tau}\tau,-j}e^{-2i\tau \phi}=S_{\bar{\tau}\tau,j}e^{-2i\tau \phi}=-S_{\tau\bar{\tau},-j}\; , \nonumber
\end{eqnarray}
where $\bar{\tau}=-\tau$.

\section{ Spinless cross sections}\label{appendb}
The amplitudes and corresponding cross sections can be obtained
by direct comparison of
the far field forms of the incoming and scattered waves in
Eq.~\ref{sj}.  For an
incoming wave front along the $x$-axis, one can write
\begin{equation}
\psi =
e^{ikr\cos\theta} \, \phi_{in}+\sum_{j} \hat{f}_{j} \, e^{ij\theta} \, \frac{e^{ikr}}{\sqrt{r}} \, \phi_{in} \, ,
\end{equation}
where $\hat{f}_{j}$ is a matrix containing the different scattering
amplitudes for each angular momentum channel $j$, and $\phi_{in}$ is a spinor representing the different
incoming valley weights. The scattering amplitudes in each
$j$-block are given by \cite{Sakurai}
\begin{equation}
\hat{f}_{j}=\frac{e^{-i\pi/4}}{\sqrt{2\pi k}}\left(
                           \begin{array}{cc}
                             (\hat{S}_{++,j}-1) & -i\hat{S}_{+-,j} \\
                            i\hat{S}_{-+,j} & (\hat{S}_{--,j}-1) \\
                           \end{array}
                         \right)
\end{equation}
and related to different cross sections by
\begin{widetext}
\begin{eqnarray}\label{crosssecs}
\sigma_{\tau\tau'}(\theta)&=&\frac{1}{2\pi
k}\left|\sum_{j=-\infty}^{\infty}{f_{\tau\tau',j}e^{ij\theta}}\right|^{2}\;,
\nonumber \\
\sigma_{t,\tau\tau'}&=&\int^{\pi}_{-\pi}{\sigma_{\tau\tau'}(\theta) \, d\theta}=\frac{1}{k}\sum_{j=-\infty}^{\infty}{\left|f_{\tau\tau',j}\right|^2}\;, \nonumber \\
\nonumber
\sigma_{tr,\tau\tau'}&=&\int^{\pi}_{-\pi}{\sigma_{\tau\tau'}(\theta)(1-\cos\theta) \, d\theta}=\sigma_{t,\tau\tau'}-\frac{1}{k}\sum_{j=-\infty}^{\infty}{Re(f_{\tau\tau',j}f^{*}_{\tau\tau',j+1})}\;,\\
\sigma_{sk,\tau\tau'}&=&\int^{\pi}_{-\pi}{\sigma_{\tau\tau'}(\theta)\sin\theta
 \, d\theta}=\frac{1}{k}\sum_{j=-\infty}^{\infty}{Im(f_{\tau\tau',j}f^{*}_{\tau\tau',j+1})} \; ,
\end{eqnarray}
\end{widetext}
where $j$ is a half integer in all the sums.  Here, $\sigma_{\tau\tau'}(\theta)$ is the
differential cross section, while $\sigma_{t,\tau\tau'}$ is the total, $\sigma_{tr,\tau\tau'}$ is the transport,
and $\sigma_{sk,\tau\tau'}$ is the skew cross section, all with valley preserving ($\tau=\tau'$) and valley flipping ($\tau\ne\tau'$) channels.
We similarly define the valley averaged quantities,
\begin{eqnarray}
&\sigma(\theta)=\sum_{\tau\tau'}\sigma_{\tau\tau'}(\theta)\; ,  \nonumber \\
&\sigma_{t}=\sum_{\tau\tau'}\sigma_{t,\tau\tau'}\;,  \nonumber \\
&\sigma_{tr}=\sum_{\tau\tau'}\sigma_{tr,\tau\tau'}\;,  \nonumber \\
&\sigma_{sk}=\sum_{\tau\tau'}\sigma_{sk,\tau\tau'}\;.
\end{eqnarray}
The presence of  $V_{T,J,h}$ perturbations ensures that the set of relations in
Eq.~\ref{eq1a} are satisfied, so that
\begin{eqnarray}\label{amplitudes}
f_{\tau\tau,j}&=&f_{\tau\tau,-j}=f_{\bar{\tau}\bar{\tau},-j}=f_{\bar{\tau}\bar{\tau},j}\; ,\\ \nonumber
f_{\tau\bar{\tau},j}&=&-f_{\bar{\tau}\tau,-j}e^{-2i\tau \phi}=f_{\bar{\tau}\tau,j}e^{-2i\tau \phi}=-f_{\tau\bar{\tau},-j}\; .
\end{eqnarray}
Consequently,
\begin{eqnarray}
&\sum_{j=-\infty}^{\infty}{f_{\tau\tau,j}e^{ij\theta}}=\sum_{j\ge\frac{1}{2}}{2f_{\tau\tau,j}\cos(j\theta)}\; , \\
\nonumber
&\sum_{j=-\infty}^{\infty}{f_{\tau\bar{\tau},j}e^{ij\theta}}=
\sum_{j\ge\frac{1}{2}}{2if_{\tau\bar{\tau},j}\sin(j\theta)}\;
, \nonumber
\end{eqnarray}
which results in Eq.~\ref{diffs} in the text. Notice also that
\begin{widetext}
\begin{eqnarray} \label{eq:B6}
&\sum_{j=-\infty}^{\infty}{f_{\tau\tau,j}f^{*}_{\tau\tau,j+1}}=%
|f_{\tau\tau,\frac{1}{2}}|^{2}+\sum_{j\ge\frac{1}{2}}{2Re(f_{\tau\tau,j}f^{*}_{\tau\tau,j+1})}\; ,  \nonumber \\
&\sum_{j=-\infty}^{\infty}{f_{\tau\bar{\tau},j}f^{*}_{\tau\bar{\tau},j+1}}=%
-|f_{\tau\bar{\tau},\frac{1}{2}}|^{2}+\sum_{j\ge\frac{1}{2}}{2Re(f_{\tau\bar{\tau},j}f^{*}_{\tau\bar{\tau},j+1})}\; ,
\end{eqnarray}
\end{widetext}
which results in
\begin{eqnarray} \label{eq:B7}
&\sum_{j=-\infty}^{\infty}{Im(f_{\tau\tau,j}f^{*}_{\tau\tau,j+1})}=0\; ,
 \nonumber \\
&\sum_{j=-\infty}^{\infty}{Im(f_{\tau\bar{\tau},j}f^{*}_{\tau\bar{\tau},j+1})}=0\;.
\end{eqnarray}

Now, when $kR\ll 1$, the only scattering channels with non vanishing
amplitudes are $j=\pm\frac{1}{2}$, and the ratio of transport to elastic
times becomes
\begin{equation}\label{ratiosoo}
\xi=\frac{\tau_{tr}}{\tau_{e}}=\frac{\sigma_{t}}{\sigma_{tr}}=\frac{\sum_{\tau\tau'}{\sigma_{t,\tau\tau'}}}{\sum_{\tau\tau'}\sigma_{tr,\tau\tau'}}\; ,\\
\end{equation}
where,
\begin{eqnarray} \label{eq:sigmas}
\sigma_{t,\tau\tau} &\approx&\frac{2}{k}|f_{\tau\tau,\frac{1}{2}}|^{2}=2\sigma_{tr,\tau\tau}\;
, \nonumber \\
\sigma_{t,\tau\bar{\tau}} &\approx&\frac{2}{k}|f_{\tau\bar{\tau},\frac{1}{2}}|^{2}\;
,\\  \nonumber
\sigma_{tr,\tau\bar{\tau}} &\approx&\frac{3}{k}|f_{\tau\bar{\tau},\frac{1}{2}}|^{2}\;
,\\  \nonumber \sigma_{sk,\tau\tau} &=&\sigma_{sk,\tau\bar{\tau}}=0\;.
\end{eqnarray}
This yields
\begin{equation}
\xi=\frac{\sigma_{t}}{\sigma_{tr}}=\frac{2\left(|f_{++,\frac{1}{2}}|^2+|f_{+-,\frac{1}{2}}|^2\right)}{|f_{++,\frac{1}{2}}|^2+3|f_{+-,\frac{1}{2}}|^2}\;
,
\end{equation}
which appears as Eq.\ \ref{eq2} in the text.

\subsection{$V_{T,J}$ perturbations}
In the presence of interactions that belong to the set $V_{T,J}$,
the conditions on the scattering matrix elements are
those resulting only from time reversal symmetry, such that they
are reduced to those in \ref{eq:SwithT},
\begin{eqnarray}\label{eqstag}
 &S_{\tau\tau,j}=S_{\bar{\tau}\bar{\tau},-j}\; ,\\ \nonumber
 &S_{\tau\bar{\tau},j}=-S_{\tau\bar{\tau},-j}\; ,\\ \nonumber
&S_{\tau\bar{\tau},j}=-S_{\bar{\tau}\tau,-j}\;  ,
\end{eqnarray}
and correspondingly,
\begin{eqnarray}\label{amplitudesstag}
&f_{\tau\tau,j}=f_{\bar{\tau}\bar{\tau},-j}\; ,\\ \nonumber
 &f_{\tau\bar{\tau},j}=-f_{\tau\bar{\tau},-j}\; ,\\ \nonumber
&f_{\tau\bar{\tau},j}=-f_{\bar{\tau}\tau,-j}  \;.
\end{eqnarray}
These relations cannot guarantee \ref{eq:B6} and \ref{eq:B7}, so that for example
$\sum^{\infty}_{j=-\infty}Im(f_{\tau\tau,j}f^{*}_{\tau\tau,j+1})$ is not identically zero by symmetry;
instead, its value depends on the details of the perturbations in the Hamiltonian.
As a consequence, one can write
\begin{eqnarray}\label{hallstag}
\sigma_{t,\tau\tau}&=&\sigma_{t,\bar{\tau}\bar{\tau}} \;, \nonumber \\
\sigma_{tr,\tau\tau}&=&\sigma_{tr,\bar{\tau}\bar{\tau}}\; , \nonumber \\
\sigma_{sk,\tau\tau}&=&-\sigma_{sk,\bar{\tau}\bar{\tau}}\;, \nonumber \\
\sigma_{t,\tau\bar{\tau}}&=&\sigma_{t,\bar{\tau}\tau}\; , \nonumber \\
\sigma_{tr,\tau\bar{\tau}}&=&\sigma_{tr,\bar{\tau}\tau}\; , \nonumber \\
\sigma_{sk,\tau\bar{\tau}}&=&\sigma_{sk,\bar{\tau}\tau}=0\; .
\end{eqnarray}

The skew parameter is defined as
\begin{equation}
\gamma=\frac{\sum_{\tau\tau'}{\sigma_{sk,\tau\tau'}}}{\sum_{\tau\tau'}{\sigma_{tr,\tau\tau'}}}
\, .
\end{equation}
The relations $\sigma_{sk,\tau\tau}\ne0$ and $\sigma_{sk,\tau\bar{\tau}}=0$ indicate
the presence of transverse valley currents, which would result in the valley Hall effect.
These currents are related to the valley skew parameter at zero temperatures by
\begin{equation}\label{valleyhall}
\gamma_{V}=\frac{j_{VH}}{j_{x}}=\frac{1}{2}(\gamma_{K}-\gamma_{K'})\;,
\end{equation}
where
\begin{equation}
\gamma_{\tau}=\frac{\sum_{\tau'}{\sigma_{sk,\tau'\tau}}}{\sum_{\tau'}{\sigma_{tr,\tau'\tau}}}\;.
\end{equation}
Moreover, for valley-polarized incident flux, $\gamma=\gamma_{V} \ne 0$ in general.
This would in principle produce a charge Hall voltage, due to the accumulation of carriers of a given
valley at one edge of the sample.

This effect is absent in the usual valley unpolarized
currents, as $\sigma_{sk,\tau\tau}=-\sigma_{sk,\bar{\tau}\bar{\tau}}$, $\sigma_{sk,\tau\tau'}=0$
and $\gamma=0$, so that there is no net charge on the edges and no Hall effect is produced.

In the case of $kR\ll 1$, we notice from Eq.~\ref{amplitudesstag} that
the amplitudes of the dominant two $j$-channels in this regime
($j=\pm\frac{1}{2}$) are not all related by symmetry, which results in $1<\xi<2$,
and dependent on the specific parameters of the Hamiltonian.
When $\Delta/|E-v|\rightarrow0$ (ultrarelativistic regime), one recovers $\xi\rightarrow2$,
while $\xi\rightarrow1$ as $\Delta/|E-v|\rightarrow1$. \cite{isotropic,resonant2}

\section{Scattering matrix and spin-orbit interactions}\label{append3}

In the presence of spin-orbit interactions that commute with the
total angular momentum, $V^{SO}_{T,J}$, and spin independent
interactions that belong to $V_{\mathcal{T}_{1}}$ (see Sec.\
\ref{sect:spin}), both
$\mathcal{T}_{1}=e^{-i\Gamma^{5}\phi}\Gamma^{5}\mathcal{A}_{2}
S_{2}\mathcal{C}$ and
$\mathcal{T}=i\mathcal{B}\Gamma^{5}\mathcal{A}_{1}S_{2}\mathcal{C}$
operators commute with $H=H_{0}+V_{\mathcal{T}_{1}}$, and allows us
relate the states with $j$ and $-j$; here one can write $\psi_j =
(J_{j-1} \, e^{-i\theta}, iJ_{j} , J_{j-1} \, e^{-i\theta} ,
-iJ_{j}, J_{j}, iJ_{j+1} \, e^{i\theta}, J_{j},$ $ -iJ_{j+1}\,
e^{i\theta} )^T e^{ij\theta}$, and where $J_{l}(kr)$ are Bessel
functions.\cite{isotropic}

In the presence of SO interactions,
an incident state could be projected into valley and spin subspaces with the aid of the corresponding
projection operators, given by
 $P_{\pm}=(1\pm\Gamma^{5})/2$ and $P_{\uparrow / \downarrow}=(1\pm S_{z})/2$.
 A general eigenstate of $H_{0}$ can then be written as
\begin{equation}
\psi=P_{+}(P_{\uparrow}+P_{\downarrow})\psi+P_{-}(P_{\uparrow}+P_{\downarrow})\psi\; .
\end{equation}
The partial wave components can be chosen to satisfy
$\mathcal{T}_{1}P_{\pm}P_{s}\psi_{j}=\pm(-1)^{j}P_{\pm}P_{\bar{s}}\psi_{-j}$ and
$\mathcal{T}P_{\pm}P_{s}\psi_{j}=\mp(-1)^{j+\frac{1}{2}}P_{\mp}P_{\bar{s}}\psi_{-j}$. Consequently
 \begin{eqnarray}
\mathcal{T}_{1}\psi_{j}&=&(-1)^{j}(P_{+}(P_{\uparrow}+P_{\downarrow})\psi_{-j}-P_{-}(P_{\uparrow}+P_{\downarrow}))\psi_{-j}\; ,  \nonumber \\
&=&(-1)^{j}\Gamma^{5}\psi_{-j}\; ,
\end{eqnarray}
 and
\begin{eqnarray}
\mathcal{T}\psi_{j}&=&(-1)^{j+\frac{1}{2}}(P_{+}(P_{\uparrow}+P_{\downarrow})\psi_{-j}-P_{-}(P_{\uparrow}+P_{\downarrow}))\psi_{-j}\; , \nonumber \\
&=&(-1)^{j+\frac{1}{2}}\Gamma^{5}\psi_{-j} \, .
\end{eqnarray}

As before, the incoming and outgoing partial waves describing the scattering problem can be written as
\begin{equation}\label{js}
\psi_{j}=\psi^{in}_{j}+S_{j}\psi^{out}_{j}\;.
\end{equation}
In this case, the scattering matrix can be written in terms of  $4\times4$, $S_{j}$-blocks, with elements $\hat{S}_{\tau\tau',ss',j}$.
By applying the operator $\mathcal{T}_{1}$ to
Eq.\ \ref{js}, we get
\begin{equation} \label{eq:C5}
\Gamma^{5}\mathcal{T}_{1}S_{j}\mathcal{T}_{1}^{-1}\Gamma^{5}=S^{\dag}_{-j} \, ,
\end{equation}
and by applying $\mathcal{T}$ we get
\begin{equation}
\Gamma^{5}\mathcal{T}S_{j}\mathcal{T}^{-1}\Gamma^{5}=S^{\dag}_{-j}\;.
\end{equation}

From these conditions and unitarity of $S_{j}$ we have,
\begin{widetext}
\begin{eqnarray}
S_{\tau\tau,ss,j}&=&S_{\bar{\tau}\bar{\tau},\bar{s}\bar{s},-j}=S_{\tau\tau,\bar{s}\bar{s},-j}\\\
\nonumber
S_{\tau\bar{\tau},ss,j}&=&-S_{\tau\bar{\tau},\bar{s}\bar{s},-j}=-S_{\bar{\tau}\tau,\bar{s}\bar{s},-j}e^{2i\tau\phi}\\
\nonumber
S_{\tau\bar{\tau},s\bar{s},j}&=&S_{\tau\bar{\tau},s\bar{s},-j}=-S_{\bar{\tau}\tau,\bar{s}s,-j}=-S_{\bar{\tau}\tau,s\bar{s},-j}e^{2i\tau\phi}=-S_{\bar{\tau}\tau,\bar{s}s,-j}e^{2i\tau\phi}\\
\nonumber
S_{\tau\tau,s\bar{s},j}&=&S_{\tau\tau,s\bar{s},-j}=-S_{\bar{\tau}\bar{\tau},\bar{s}s,-j}e^{2i\tau\phi} \, .\\
\nonumber
\end{eqnarray}
\end{widetext}
The scattering amplitudes are given by
\begin{widetext}
\begin{equation}
\hat{f}_j(\theta)=\frac{e^{-i\pi/4}}{\sqrt{2\pi k}}\left(
  \begin{array}{cccc}
    (\hat{S}_{++\uparrow\uparrow,j}-1) & -i\hat{S}_{+-,\uparrow\uparrow,j} & -i\hat{S}_{++,\uparrow\downarrow,j} & -\hat{S}_{+-,\uparrow\downarrow,j} \\
    i\hat{S}_{-+,\uparrow\uparrow,j} & (\hat{S}_{--\uparrow\uparrow,j}-1) & \hat{S}_{-+,\uparrow\downarrow,j} &  -i\hat{S}_{--,\uparrow\downarrow,j} \\
    i\hat{S}_{++,\downarrow\uparrow,j} & \hat{S}_{+-,\downarrow\uparrow,j} & (\hat{S}_{++\downarrow\downarrow,j}-1) & -i\hat{S}_{+-,\downarrow\downarrow,j}  \\
    -\hat{S}_{-+,\downarrow\uparrow,j} & i\hat{S}_{--,\downarrow\uparrow,j} & i\hat{S}_{-+,\downarrow\downarrow,j} & (\hat{S}_{--\downarrow\downarrow,j}-1) \\
  \end{array}
\right)\; ,
\end{equation}
and the different cross sections
\begin{eqnarray}
\sigma_{\tau\tau',ss'}(\theta)&=&\frac{1}{2\pi
k}\left|\sum_{j=-\infty}^{\infty}{f_{j,\tau\tau',ss'}e^{ij\theta}}\right|^{2}\;,\\
\nonumber
\sigma_{t,\tau\tau',ss'}&=&\frac{1}{k}\sum_{j=-\infty}^{\infty}{\left|f_{j,\tau\tau',ss'}\right|^2}\;,\\
\nonumber
\sigma_{tr,\tau\tau',ss'}&=&\sigma_{t,\tau\tau',ss'}-\frac{1}{k}\sum_{j=-\infty}^{\infty}{Re(f_{j,\tau\tau',ss'}f^{*}_{j+1\tau\tau',ss'})}\;,\\
\nonumber
\sigma_{sk,\tau\tau',ss'}&=&\frac{1}{k}\sum_{j=-\infty}^{\infty}{Im(f_{j,\tau\tau',ss'}f^{*}_{j+1\tau\tau',ss'})} \, .
\end{eqnarray}
\end{widetext}
These help define the ratio of transport to elastic times in the presence of SOI,
\begin{equation}
\xi_{SO}=\frac{\sum_{\tau\tau',ss'}{\sigma_{t,\tau\tau',ss'}}}{\sum_{\tau\tau',ss'}{\sigma_{tr,\tau\tau',ss'}}}\;,
\end{equation}
and skew parameter
\begin{eqnarray}\label{skewparmater}
\gamma_{S}&=&\frac{1}{2}(\gamma_{\uparrow}-\gamma_{\downarrow})\; ,\\
\nonumber \gamma_{s}&=&\frac{\sum_{\tau,\tau',s'}{\sigma_{sk,\tau\tau',s's}}}{\sum_{\tau,\tau',s'}{\sigma_{tr,\tau\tau',s's}}}\;.
\end{eqnarray}

For interactions commuting with $\mathcal{T}$ and
$\mathcal{T}_{1}$, we have the relations
\begin{widetext}
\begin{eqnarray}\label{spinrelations}
\sum^{\infty}_{j=-\infty}{f_{\tau\tau,ss,j}f^{*}_{\tau\tau,ss,j+1}}&=&(\sum^{\infty}_{j=-\infty}{f_{\bar{\tau}\bar{\tau},\bar{s}\bar{s},j}f^{*}_{\bar{\tau}\bar{\tau},\bar{s}\bar{s},j+1}})^{*}=(\sum^{\infty}_{j=-\infty}{f_{\tau\tau,\bar{s}\bar{s},j}f^{*}_{\tau\tau,\bar{s}\bar{s},j+1}})^{*}=\sum^{\infty}_{j=-\infty}{f_{\bar{\tau}\bar{\tau},ss,j}f^{*}_{\bar{\tau}\bar{\tau},ss,j+1}}\;
, \nonumber \\ \nonumber
\sum^{\infty}_{j=-\infty}{f_{\tau\bar{\tau},ss,j}f^{*}_{\tau\bar{\tau},ss,j+1}}&=&\sum^{\infty}_{j=-\infty}{f_{\bar{\tau}\tau,ss,j}f^{*}_{\bar{\tau}\tau,ss,j+1}}=(\sum^{\infty}_{j=-\infty}{f_{\bar{\tau}\tau,\bar{s}\bar{s},j}f^{*}_{\bar{\tau}\tau,\bar{s}\bar{s},j+1}})^{*}=(\sum^{\infty}_{j=-\infty}{f_{\tau\bar{\tau},\bar{s}\bar{s},j}f^{*}_{\tau\bar{\tau},\bar{s}\bar{s},j+1}})^{*}\;
,\\
\sum^{\infty}_{j=-\infty}{Im(f_{\tau\tau',s\bar{s},j}f^{*}_{\tau\tau',s\bar{s},j+1})}&=&\sum^{\infty}_{j=-\infty}{Im(f_{\tau\tau',\bar{s}s,j}f^{*}_{\tau\tau',\bar{s}s,j+1})}=0\; ,
\end{eqnarray}
\end{widetext}
and consequently,
\begin{eqnarray} \label{eq:C13}
\sigma_{t,\tau\tau,ss}&=&\sigma_{t,\tau\tau,\bar{s}\bar{s}}=\sigma_{t,\bar{\tau}\bar{\tau},\bar{s}\bar{s}}=\sigma_{t,\bar{\tau}\bar{\tau},ss}\;
,\\ \nonumber
\sigma_{tr,\tau\tau,ss}&=&\sigma_{tr,\tau\tau,\bar{s}\bar{s}}=\sigma_{tr,\bar{\tau}\bar{\tau},\bar{s}\bar{s}}=\sigma_{tr,\bar{\tau}\bar{\tau},ss}\;
,\\ \nonumber
\sigma_{sk,\tau\tau,ss}&=&-\sigma_{sk,\tau\tau,\bar{s}\bar{s}}=-\sigma_{sk,\bar{\tau}\bar{\tau},\bar{s}\bar{s}}=\sigma_{sk,\bar{\tau}\bar{\tau},ss}\;
,\\ \nonumber
\sigma_{t,\tau\bar{\tau},ss}&=&\sigma_{t,\bar{\tau}\tau,ss}=\sigma_{t,\bar{\tau}\tau,\bar{s}\bar{s}}=\sigma_{t,\tau\bar{\tau},\bar{s}\bar{s}}\;
,\\ \nonumber
\sigma_{tr,\tau\bar{\tau},ss}&=&\sigma_{tr,\bar{\tau}\tau,ss}=\sigma_{tr,\bar{\tau}\tau,\bar{s}\bar{s}}=\sigma_{tr,\tau\bar{\tau},\bar{s}\bar{s}}\;
,\\ \nonumber
\sigma_{sk,\tau\bar{\tau},ss}&=&\sigma_{sk,\bar{\tau}\tau,ss}=-\sigma_{sk,\bar{\tau}\tau,\bar{s}\bar{s}}=-\sigma_{sk,\tau\bar{\tau},\bar{s}\bar{s}}\;
,\\ \nonumber
\sigma_{t,\tau\tau',s\bar{s}}&=&\sigma_{t,\tau\tau',\bar{s}s}\; ,\\
 \nonumber
\sigma_{tr,\tau\tau',s\bar{s}}&=&\sigma_{tr,\tau\tau',\bar{s}s}\; ,\\
\nonumber
\sigma_{sk,\tau\tau',s\bar{s}}&=&\sigma_{sk,\tau\tau',\bar{s}s}=0\;
.\\ \nonumber
\end{eqnarray}

This leads to $\gamma_{\uparrow}=-\gamma_{\downarrow}$ and
$\gamma_S=\gamma_{\uparrow}$, so that even for a valley
and spin unpolarized incident wave, there is accumulation of
different spins on edges of the sample, and the corresponding appearance of the spin Hall effect.

In the presence of SOIs the total angular momentum is an
integer ($j\in Z$), as noted before. Then for $kR\ll 1$, there are {\em three} dominant angular momentum
channels in the scattering process, $j=0,\pm1$. In the case
of resonance, these channels resonate at different energies,
leading to a drop of $\xi_{SO}$ towards $1$ at these
energies, as the scattering process is predominantly caused by only one
of the channels for a given spin kind.

In the case where the perturbations present {\em do not} commute with
$\mathcal{T}_{1}$, instead of \ref{eq:C13} we have
\begin{eqnarray}\label{skewspinsttagered}
\sigma_{t,\tau\tau,ss}&=&\sigma_{t,\bar{\tau}\bar{\tau},\bar{s}\bar{s}}\;,\\
\nonumber
\sigma_{tr,\tau\tau,ss}&=&\sigma_{tr,\bar{\tau}\bar{\tau},\bar{s}\bar{s}}\;,\\
\nonumber
\sigma_{sk,\tau\tau,ss}&=&-\sigma_{sk,\bar{\tau}\bar{\tau},\bar{s}\bar{s}}\;,\\
\nonumber
\sigma_{t,\tau\bar{\tau},ss}&=&\sigma_{t,\tau\bar{\tau},\bar{s}\bar{s}}\;
,\\ \nonumber
\sigma_{tr,\tau\bar{\tau},ss}&=&\sigma_{tr,\tau\bar{\tau},\bar{s}\bar{s}}\;
,\\ \nonumber
\sigma_{sk,\tau\bar{\tau},ss}&=&-\sigma_{sk,\tau\bar{\tau},\bar{s}\bar{s}}\;
,\\ \nonumber
\sigma_{sk,\tau\tau',s\bar{s}}&=&\sigma_{sk,\tau\tau',\bar{s}s}=0\; . \nonumber
\end{eqnarray}

In this case, $\gamma_{S}$ gets reduced from its optimal value, which
indicates that $V_{\mathcal{T},J}$ perturbations that do no commute with $\mathcal{T}_{1}$ lead to a reduction in 
the possible SHE: the competition with valley skew scattering (produced by a staggered perturbation, for example) 
will result in a reduction of the spin imbalance accumulated at the edges 
of the samples.  As such  $V_{\mathcal{T},J}$ perturbations
increase, $\gamma_{S}$ may even become negligible, as $\gamma_{\uparrow} \simeq \gamma_{\downarrow}$ 
if valley skewness dominates, as schematically represented in Fig.\ \ref{scem}.

\bibliography{references}

\begin{thebibliography}{59}%
\makeatletter
\providecommand \@ifxundefined [1]{%
 \@ifx{#1\undefined}
}%
\providecommand \@ifnum [1]{%
 \ifnum #1\expandafter \@firstoftwo
 \else \expandafter \@secondoftwo
 \fi
}%
\providecommand \@ifx [1]{%
 \ifx #1\expandafter \@firstoftwo
 \else \expandafter \@secondoftwo
 \fi
}%
\providecommand \natexlab [1]{#1}%
\providecommand \enquote  [1]{``#1''}%
\providecommand \bibnamefont  [1]{#1}%
\providecommand \bibfnamefont [1]{#1}%
\providecommand \citenamefont [1]{#1}%
\providecommand \href@noop [0]{\@secondoftwo}%
\providecommand \href [0]{\begingroup \@sanitize@url \@href}%
\providecommand \@href[1]{\@@startlink{#1}\@@href}%
\providecommand \@@href[1]{\endgroup#1\@@endlink}%
\providecommand \@sanitize@url [0]{\catcode `\\12\catcode `\$12\catcode
  `\&12\catcode `\#12\catcode `\^12\catcode `\_12\catcode `\%12\relax}%
\providecommand \@@startlink[1]{}%
\providecommand \@@endlink[0]{}%
\providecommand \url  [0]{\begingroup\@sanitize@url \@url }%
\providecommand \@url [1]{\endgroup\@href {#1}{\urlprefix }}%
\providecommand \urlprefix  [0]{URL }%
\providecommand \Eprint [0]{\href }%
\providecommand \doibase [0]{http://dx.doi.org/}%
\providecommand \selectlanguage [0]{\@gobble}%
\providecommand \bibinfo  [0]{\@secondoftwo}%
\providecommand \bibfield  [0]{\@secondoftwo}%
\providecommand \translation [1]{[#1]}%
\providecommand \BibitemOpen [0]{}%
\providecommand \bibitemStop [0]{}%
\providecommand \bibitemNoStop [0]{.\EOS\space}%
\providecommand \EOS [0]{\spacefactor3000\relax}%
\providecommand \BibitemShut  [1]{\csname bibitem#1\endcsname}%
\let\auto@bib@innerbib\@empty
\bibitem [{\citenamefont {Castro~Neto}\ \emph {et~al.}(2009)\citenamefont
  {Castro~Neto}, \citenamefont {Guinea}, \citenamefont {Peres}, \citenamefont
  {Novoselov},\ and\ \citenamefont {Geim}}]{elctronincproperties}%
  \BibitemOpen
  \bibfield  {author} {\bibinfo {author} {\bibfnamefont {A.~H.}\ \bibnamefont
  {Castro~Neto}}, \bibinfo {author} {\bibfnamefont {F.}~\bibnamefont {Guinea}},
  \bibinfo {author} {\bibfnamefont {N.~M.~R.}\ \bibnamefont {Peres}}, \bibinfo
  {author} {\bibfnamefont {K.~S.}\ \bibnamefont {Novoselov}}, \ and\ \bibinfo
  {author} {\bibfnamefont {A.~K.}\ \bibnamefont {Geim}},\ }\href {\doibase
  10.1103/RevModPhys.81.109} {\bibfield  {journal} {\bibinfo  {journal} {Rev.
  Mod. Phys.}\ }\textbf {\bibinfo {volume} {81}},\ \bibinfo {pages} {109}
  (\bibinfo {year} {2009})}\BibitemShut {NoStop}%
\bibitem [{\citenamefont {Kane}\ and\ \citenamefont {Mele}(2005)}]{quntums}%
  \BibitemOpen
  \bibfield  {author} {\bibinfo {author} {\bibfnamefont {C.~L.}\ \bibnamefont
  {Kane}}\ and\ \bibinfo {author} {\bibfnamefont {E.~J.}\ \bibnamefont
  {Mele}},\ }\href {\doibase 10.1103/PhysRevLett.95.226801} {\bibfield
  {journal} {\bibinfo  {journal} {Phys. Rev. Lett.}\ }\textbf {\bibinfo
  {volume} {95}},\ \bibinfo {pages} {226801} (\bibinfo {year}
  {2005})}\BibitemShut {NoStop}%
\bibitem [{\citenamefont {Fu}\ and\ \citenamefont {Kane}(2006)}]{timerev}%
  \BibitemOpen
  \bibfield  {author} {\bibinfo {author} {\bibfnamefont {L.}~\bibnamefont
  {Fu}}\ and\ \bibinfo {author} {\bibfnamefont {C.~L.}\ \bibnamefont {Kane}},\
  }\href {\doibase 10.1103/PhysRevB.74.195312} {\bibfield  {journal} {\bibinfo
  {journal} {Phys. Rev. B}\ }\textbf {\bibinfo {volume} {74}},\ \bibinfo
  {pages} {195312} (\bibinfo {year} {2006})}\BibitemShut {NoStop}%
\bibitem [{\citenamefont {Hsieh}\ \emph {et~al.}(2008)\citenamefont {Hsieh},
  \citenamefont {Qian}, \citenamefont {Wray}, \citenamefont {Xia},
  \citenamefont {Hor}, \citenamefont {Cava},\ and\ \citenamefont
  {Hasan}}]{topoexp1}%
  \BibitemOpen
  \bibfield  {author} {\bibinfo {author} {\bibfnamefont {D.}~\bibnamefont
  {Hsieh}}, \bibinfo {author} {\bibfnamefont {D.}~\bibnamefont {Qian}},
  \bibinfo {author} {\bibfnamefont {L.}~\bibnamefont {Wray}}, \bibinfo {author}
  {\bibfnamefont {Y.}~\bibnamefont {Xia}}, \bibinfo {author} {\bibfnamefont
  {Y.~S.}\ \bibnamefont {Hor}}, \bibinfo {author} {\bibfnamefont {R.~J.}\
  \bibnamefont {Cava}}, \ and\ \bibinfo {author} {\bibfnamefont {M.~Z.}\
  \bibnamefont {Hasan}},\ }\href {\doibase 10.1038/nature06843} {\bibfield
  {journal} {\bibinfo  {journal} {Nature}\ }\textbf {\bibinfo {volume} {452}},\
  \bibinfo {pages} {970} (\bibinfo {year} {2008})}\BibitemShut {NoStop}%
\bibitem [{\citenamefont {Schnyder}\ \emph {et~al.}(2008)\citenamefont
  {Schnyder}, \citenamefont {Ryu}, \citenamefont {Furusaki},\ and\
  \citenamefont {Ludwig}}]{topol3}%
  \BibitemOpen
  \bibfield  {author} {\bibinfo {author} {\bibfnamefont {A.~P.}\ \bibnamefont
  {Schnyder}}, \bibinfo {author} {\bibfnamefont {S.}~\bibnamefont {Ryu}},
  \bibinfo {author} {\bibfnamefont {A.}~\bibnamefont {Furusaki}}, \ and\
  \bibinfo {author} {\bibfnamefont {A.~W.~W.}\ \bibnamefont {Ludwig}},\ }\href
  {\doibase 10.1103/PhysRevB.78.195125} {\bibfield  {journal} {\bibinfo
  {journal} {Phys. Rev. B}\ }\textbf {\bibinfo {volume} {78}},\ \bibinfo
  {pages} {195125} (\bibinfo {year} {2008})}\BibitemShut {NoStop}%
\bibitem [{\citenamefont {Hasan}\ and\ \citenamefont {Kane}(2010)}]{topol2}%
  \BibitemOpen
  \bibfield  {author} {\bibinfo {author} {\bibfnamefont {M.~Z.}\ \bibnamefont
  {Hasan}}\ and\ \bibinfo {author} {\bibfnamefont {C.~L.}\ \bibnamefont
  {Kane}},\ }\href {\doibase 10.1103/RevModPhys.82.3045} {\bibfield  {journal}
  {\bibinfo  {journal} {Rev. Mod. Phys.}\ }\textbf {\bibinfo {volume} {82}},\
  \bibinfo {pages} {3045} (\bibinfo {year} {2010})}\BibitemShut {NoStop}%
\bibitem [{\citenamefont {Qi}\ and\ \citenamefont {Zhang}(2011)}]{topol1}%
  \BibitemOpen
  \bibfield  {author} {\bibinfo {author} {\bibfnamefont {X.-L.}\ \bibnamefont
  {Qi}}\ and\ \bibinfo {author} {\bibfnamefont {S.-C.}\ \bibnamefont {Zhang}},\
  }\href {\doibase 10.1103/RevModPhys.83.1057} {\bibfield  {journal} {\bibinfo
  {journal} {Rev. Mod. Phys.}\ }\textbf {\bibinfo {volume} {83}},\ \bibinfo
  {pages} {1057} (\bibinfo {year} {2011})}\BibitemShut {NoStop}%
\bibitem [{\citenamefont {Castro~Neto}\ and\ \citenamefont
  {Guinea}(2009)}]{impurityso}%
  \BibitemOpen
  \bibfield  {author} {\bibinfo {author} {\bibfnamefont {A.~H.}\ \bibnamefont
  {Castro~Neto}}\ and\ \bibinfo {author} {\bibfnamefont {F.}~\bibnamefont
  {Guinea}},\ }\href {\doibase 10.1103/PhysRevLett.103.026804} {\bibfield
  {journal} {\bibinfo  {journal} {Phys. Rev. Lett.}\ }\textbf {\bibinfo
  {volume} {103}},\ \bibinfo {pages} {026804} (\bibinfo {year}
  {2009})}\BibitemShut {NoStop}%
\bibitem [{\citenamefont {Balakrishnan}\ \emph {et~al.}(2013)\citenamefont
  {Balakrishnan}, \citenamefont {Kok Wai~Koon}, \citenamefont {Jaiswal},
  \citenamefont {Castro~Neto},\ and\ \citenamefont {\"Ozyilmaz}}]{colossal}%
  \BibitemOpen
  \bibfield  {author} {\bibinfo {author} {\bibfnamefont {J.}~\bibnamefont
  {Balakrishnan}}, \bibinfo {author} {\bibfnamefont {G.}~\bibnamefont {Kok
  Wai~Koon}}, \bibinfo {author} {\bibfnamefont {M.}~\bibnamefont {Jaiswal}},
  \bibinfo {author} {\bibfnamefont {A.~H.}\ \bibnamefont {Castro~Neto}}, \ and\
  \bibinfo {author} {\bibfnamefont {B.}~\bibnamefont {\"Ozyilmaz}},\ }\href
  {\doibase 10.1038/nphys2576} {\bibfield  {journal} {\bibinfo  {journal} {Nat.
  Phys.}\ }\textbf {\bibinfo {volume} {9}},\ \bibinfo {pages} {284} (\bibinfo
  {year} {2013})}\BibitemShut {NoStop}%
\bibitem [{\citenamefont {Ferreira}\ \emph {et~al.}(2014)\citenamefont
  {Ferreira}, \citenamefont {Rappoport}, \citenamefont {Cazalilla},\ and\
  \citenamefont {Castro~Neto}}]{ferreira1}%
  \BibitemOpen
  \bibfield  {author} {\bibinfo {author} {\bibfnamefont {A.}~\bibnamefont
  {Ferreira}}, \bibinfo {author} {\bibfnamefont {T.~G.}\ \bibnamefont
  {Rappoport}}, \bibinfo {author} {\bibfnamefont {M.~A.}\ \bibnamefont
  {Cazalilla}}, \ and\ \bibinfo {author} {\bibfnamefont {A.~H.}\ \bibnamefont
  {Castro~Neto}},\ }\href@noop {} {\bibfield  {journal} {\bibinfo  {journal}
  {Phys. Rev. Lett.}\ }\textbf {\bibinfo {volume} {112}},\ \bibinfo {pages}
  {066601} (\bibinfo {year} {2014})}\BibitemShut {NoStop}%
\bibitem [{\citenamefont {Balakrishnan}\ \emph {et~al.}(2014)\citenamefont
  {Balakrishnan}, \citenamefont {Koon}, \citenamefont {Avsar}, \citenamefont
  {Ho}, \citenamefont {Lee}, \citenamefont {Jaiswal}, \citenamefont {Baeck},
  \citenamefont {Ahn}, \citenamefont {Ferreira}, \citenamefont {Cazalilla},
  \citenamefont {Neto},\ and\ \citenamefont {\"Ozyilmaz}}]{nanoparticles}%
  \BibitemOpen
  \bibfield  {author} {\bibinfo {author} {\bibfnamefont {J.}~\bibnamefont
  {Balakrishnan}}, \bibinfo {author} {\bibfnamefont {G.~K.~W.}\ \bibnamefont
  {Koon}}, \bibinfo {author} {\bibfnamefont {A.}~\bibnamefont {Avsar}},
  \bibinfo {author} {\bibfnamefont {Y.}~\bibnamefont {Ho}}, \bibinfo {author}
  {\bibfnamefont {J.~H.}\ \bibnamefont {Lee}}, \bibinfo {author} {\bibfnamefont
  {M.}~\bibnamefont {Jaiswal}}, \bibinfo {author} {\bibfnamefont {S.-J.}\
  \bibnamefont {Baeck}}, \bibinfo {author} {\bibfnamefont {J.-H.}\ \bibnamefont
  {Ahn}}, \bibinfo {author} {\bibfnamefont {A.}~\bibnamefont {Ferreira}},
  \bibinfo {author} {\bibfnamefont {M.~A.}\ \bibnamefont {Cazalilla}}, \bibinfo
  {author} {\bibfnamefont {A.~H.~C.}\ \bibnamefont {Neto}}, \ and\ \bibinfo
  {author} {\bibfnamefont {B.}~\bibnamefont {\"Ozyilmaz}},\ }\href@noop {}
  {\bibfield  {journal} {\bibinfo  {journal} {Nat. Commun.}\ }\textbf {\bibinfo
  {volume} {5}} (\bibinfo {year} {2014})}\BibitemShut {NoStop}%
\bibitem [{\citenamefont {Novoselov}\ \emph {et~al.}(2005)\citenamefont
  {Novoselov}, \citenamefont {Jiang}, \citenamefont {Schedin}, \citenamefont
  {Booth}, \citenamefont {Khotkevich}, \citenamefont {Morozov},\ and\
  \citenamefont {Geim}}]{layered}%
  \BibitemOpen
  \bibfield  {author} {\bibinfo {author} {\bibfnamefont {K.~S.}\ \bibnamefont
  {Novoselov}}, \bibinfo {author} {\bibfnamefont {D.}~\bibnamefont {Jiang}},
  \bibinfo {author} {\bibfnamefont {F.}~\bibnamefont {Schedin}}, \bibinfo
  {author} {\bibfnamefont {T.~J.}\ \bibnamefont {Booth}}, \bibinfo {author}
  {\bibfnamefont {V.~V.}\ \bibnamefont {Khotkevich}}, \bibinfo {author}
  {\bibfnamefont {S.~V.}\ \bibnamefont {Morozov}}, \ and\ \bibinfo {author}
  {\bibfnamefont {A.~K.}\ \bibnamefont {Geim}},\ }\href {\doibase
  10.1073/pnas.0502848102} {\bibfield  {journal} {\bibinfo  {journal} {PNAS}\
  }\textbf {\bibinfo {volume} {102}},\ \bibinfo {pages} {10451} (\bibinfo
  {year} {2005})}\BibitemShut {NoStop}%
\bibitem [{\citenamefont {Novoselov}\ and\ \citenamefont
  {Neto}(2012)}]{layered1}%
  \BibitemOpen
  \bibfield  {author} {\bibinfo {author} {\bibfnamefont {K.~S.}\ \bibnamefont
  {Novoselov}}\ and\ \bibinfo {author} {\bibfnamefont {A.~H.~C.}\ \bibnamefont
  {Neto}},\ }\href@noop {} {\bibfield  {journal} {\bibinfo  {journal} {Phys.
  Scripta}\ }\textbf {\bibinfo {volume} {2012}},\ \bibinfo {pages} {014006}
  (\bibinfo {year} {2012})}\BibitemShut {NoStop}%
\bibitem [{\citenamefont {Varykhalov}\ \emph {et~al.}(2008)\citenamefont
  {Varykhalov}, \citenamefont {S\'anchez-Barriga}, \citenamefont {Shikin},
  \citenamefont {Biswas}, \citenamefont {Vescovo}, \citenamefont {Rybkin},
  \citenamefont {Marchenko},\ and\ \citenamefont {Rader}}]{Ni1}%
  \BibitemOpen
  \bibfield  {author} {\bibinfo {author} {\bibfnamefont {A.}~\bibnamefont
  {Varykhalov}}, \bibinfo {author} {\bibfnamefont {J.}~\bibnamefont
  {S\'anchez-Barriga}}, \bibinfo {author} {\bibfnamefont {A.~M.}\ \bibnamefont
  {Shikin}}, \bibinfo {author} {\bibfnamefont {C.}~\bibnamefont {Biswas}},
  \bibinfo {author} {\bibfnamefont {E.}~\bibnamefont {Vescovo}}, \bibinfo
  {author} {\bibfnamefont {A.}~\bibnamefont {Rybkin}}, \bibinfo {author}
  {\bibfnamefont {D.}~\bibnamefont {Marchenko}}, \ and\ \bibinfo {author}
  {\bibfnamefont {O.}~\bibnamefont {Rader}},\ }\href {\doibase
  10.1103/PhysRevLett.101.157601} {\bibfield  {journal} {\bibinfo  {journal}
  {Phys. Rev. Lett.}\ }\textbf {\bibinfo {volume} {101}},\ \bibinfo {pages}
  {157601} (\bibinfo {year} {2008})}\BibitemShut {NoStop}%
\bibitem [{\citenamefont {Pi}\ \emph {et~al.}(2010)\citenamefont {Pi},
  \citenamefont {Han}, \citenamefont {McCreary}, \citenamefont {Swartz},
  \citenamefont {Li},\ and\ \citenamefont {Kawakami}}]{golddep}%
  \BibitemOpen
  \bibfield  {author} {\bibinfo {author} {\bibfnamefont {K.}~\bibnamefont
  {Pi}}, \bibinfo {author} {\bibfnamefont {W.}~\bibnamefont {Han}}, \bibinfo
  {author} {\bibfnamefont {K.~M.}\ \bibnamefont {McCreary}}, \bibinfo {author}
  {\bibfnamefont {A.~G.}\ \bibnamefont {Swartz}}, \bibinfo {author}
  {\bibfnamefont {Y.}~\bibnamefont {Li}}, \ and\ \bibinfo {author}
  {\bibfnamefont {R.~K.}\ \bibnamefont {Kawakami}},\ }\href {\doibase
  10.1103/PhysRevLett.104.187201} {\bibfield  {journal} {\bibinfo  {journal}
  {Phys. Rev. Lett.}\ }\textbf {\bibinfo {volume} {104}},\ \bibinfo {pages}
  {187201} (\bibinfo {year} {2010})}\BibitemShut {NoStop}%
\bibitem [{\citenamefont {Marchenko}\ \emph {et~al.}(2012)\citenamefont
  {Marchenko}, \citenamefont {Varykhalov}, \citenamefont {Scholz},
  \citenamefont {Bihlmayer}, \citenamefont {Rashba}, \citenamefont {Rybkin},
  \citenamefont {Shikin},\ and\ \citenamefont {Rader}}]{Nii}%
  \BibitemOpen
  \bibfield  {author} {\bibinfo {author} {\bibfnamefont {D.}~\bibnamefont
  {Marchenko}}, \bibinfo {author} {\bibfnamefont {A.}~\bibnamefont
  {Varykhalov}}, \bibinfo {author} {\bibfnamefont {M.}~\bibnamefont {Scholz}},
  \bibinfo {author} {\bibfnamefont {G.}~\bibnamefont {Bihlmayer}}, \bibinfo
  {author} {\bibfnamefont {E.}~\bibnamefont {Rashba}}, \bibinfo {author}
  {\bibfnamefont {A.}~\bibnamefont {Rybkin}}, \bibinfo {author} {\bibfnamefont
  {A.}~\bibnamefont {Shikin}}, \ and\ \bibinfo {author} {\bibfnamefont
  {O.}~\bibnamefont {Rader}},\ }\href {\doibase 10.1038/ncomms2227} {\bibfield
  {journal} {\bibinfo  {journal} {Nat. Commun.}\ }\textbf {\bibinfo {volume}
  {3}},\ \bibinfo {pages} {1232} (\bibinfo {year} {2012})}\BibitemShut
  {NoStop}%
\bibitem [{\citenamefont {Zhang}\ \emph {et~al.}(2014)\citenamefont {Zhang},
  \citenamefont {Triola},\ and\ \citenamefont {Rossi}}]{rossi}%
  \BibitemOpen
  \bibfield  {author} {\bibinfo {author} {\bibfnamefont {J.}~\bibnamefont
  {Zhang}}, \bibinfo {author} {\bibfnamefont {C.}~\bibnamefont {Triola}}, \
  and\ \bibinfo {author} {\bibfnamefont {E.}~\bibnamefont {Rossi}},\
  }\href@noop {} {\bibfield  {journal} {\bibinfo  {journal} {Phys. Rev. Lett.}\
  }\textbf {\bibinfo {volume} {112}},\ \bibinfo {pages} {096802} (\bibinfo
  {year} {2014})}\BibitemShut {NoStop}%
\bibitem [{\citenamefont {Woods}\ \emph {et~al.}(2014)\citenamefont {Woods},
  \citenamefont {Britnell}, \citenamefont {Eckmann}, \citenamefont {Ma},
  \citenamefont {Lu}, \citenamefont {Guo}, \citenamefont {Lin}, \citenamefont
  {Yu}, \citenamefont {Cao}, \citenamefont {Gorbachev}, \citenamefont
  {Kretinin}, \citenamefont {Park}, \citenamefont {Ponomarenko}, \citenamefont
  {Katsnelson}, \citenamefont {Gornostyrev}, \citenamefont {Watanabe},
  \citenamefont {Taniguchi}, \citenamefont {Casiraghi}, \citenamefont {Gao},
  \citenamefont {Geim},\ and\ \citenamefont {Novoselov}}]{moire}%
  \BibitemOpen
  \bibfield  {author} {\bibinfo {author} {\bibfnamefont {C.~R.}\ \bibnamefont
  {Woods}}, \bibinfo {author} {\bibfnamefont {L.}~\bibnamefont {Britnell}},
  \bibinfo {author} {\bibfnamefont {A.}~\bibnamefont {Eckmann}}, \bibinfo
  {author} {\bibfnamefont {R.~S.}\ \bibnamefont {Ma}}, \bibinfo {author}
  {\bibfnamefont {J.~C.}\ \bibnamefont {Lu}}, \bibinfo {author} {\bibfnamefont
  {H.~M.}\ \bibnamefont {Guo}}, \bibinfo {author} {\bibfnamefont
  {X.}~\bibnamefont {Lin}}, \bibinfo {author} {\bibfnamefont {G.~L.}\
  \bibnamefont {Yu}}, \bibinfo {author} {\bibfnamefont {Y.}~\bibnamefont
  {Cao}}, \bibinfo {author} {\bibfnamefont {R.~V.}\ \bibnamefont {Gorbachev}},
  \bibinfo {author} {\bibfnamefont {A.~V.}\ \bibnamefont {Kretinin}}, \bibinfo
  {author} {\bibfnamefont {J.}~\bibnamefont {Park}}, \bibinfo {author}
  {\bibfnamefont {L.~A.}\ \bibnamefont {Ponomarenko}}, \bibinfo {author}
  {\bibfnamefont {M.~I.}\ \bibnamefont {Katsnelson}}, \bibinfo {author}
  {\bibfnamefont {Y.~N.}\ \bibnamefont {Gornostyrev}}, \bibinfo {author}
  {\bibfnamefont {K.}~\bibnamefont {Watanabe}}, \bibinfo {author}
  {\bibfnamefont {T.}~\bibnamefont {Taniguchi}}, \bibinfo {author}
  {\bibfnamefont {C.}~\bibnamefont {Casiraghi}}, \bibinfo {author}
  {\bibfnamefont {H.-J.}\ \bibnamefont {Gao}}, \bibinfo {author} {\bibfnamefont
  {A.~K.}\ \bibnamefont {Geim}}, \ and\ \bibinfo {author} {\bibfnamefont
  {K.~S.}\ \bibnamefont {Novoselov}},\ }\href@noop {} {\bibfield  {journal}
  {\bibinfo  {journal} {Nat. Phys.}\ }\textbf {\bibinfo {volume} {10}},\
  \bibinfo {pages} {451} (\bibinfo {year} {2014})}\BibitemShut {NoStop}%
\bibitem [{\citenamefont {Ando}(2000)}]{curvAndo}%
  \BibitemOpen
  \bibfield  {author} {\bibinfo {author} {\bibfnamefont {T.}~\bibnamefont
  {Ando}},\ }\href {\doibase 10.1143/JPSJ.69.1757} {\bibfield  {journal}
  {\bibinfo  {journal} {J. Phys. Soc. Jap.}\ }\textbf {\bibinfo {volume}
  {69}},\ \bibinfo {pages} {1757} (\bibinfo {year} {2000})}\BibitemShut
  {NoStop}%
\bibitem [{\citenamefont {De~Martino}\ \emph {et~al.}(2002)\citenamefont
  {De~Martino}, \citenamefont {Egger}, \citenamefont {Hallberg},\ and\
  \citenamefont {Balseiro}}]{newcurv1}%
  \BibitemOpen
  \bibfield  {author} {\bibinfo {author} {\bibfnamefont {A.}~\bibnamefont
  {De~Martino}}, \bibinfo {author} {\bibfnamefont {R.}~\bibnamefont {Egger}},
  \bibinfo {author} {\bibfnamefont {K.}~\bibnamefont {Hallberg}}, \ and\
  \bibinfo {author} {\bibfnamefont {C.~A.}\ \bibnamefont {Balseiro}},\ }\href
  {\doibase 10.1103/PhysRevLett.88.206402} {\bibfield  {journal} {\bibinfo
  {journal} {Phys. Rev. Lett.}\ }\textbf {\bibinfo {volume} {88}},\ \bibinfo
  {pages} {206402} (\bibinfo {year} {2002})}\BibitemShut {NoStop}%
\bibitem [{\citenamefont {Chico}\ \emph {et~al.}(2004)\citenamefont {Chico},
  \citenamefont {L\'opez-Sancho},\ and\ \citenamefont {Mu\~noz}}]{newcurv2}%
  \BibitemOpen
  \bibfield  {author} {\bibinfo {author} {\bibfnamefont {L.}~\bibnamefont
  {Chico}}, \bibinfo {author} {\bibfnamefont {M.~P.}\ \bibnamefont
  {L\'opez-Sancho}}, \ and\ \bibinfo {author} {\bibfnamefont {M.~C.}\
  \bibnamefont {Mu\~noz}},\ }\href {\doibase 10.1103/PhysRevLett.93.176402}
  {\bibfield  {journal} {\bibinfo  {journal} {Phys. Rev. Lett.}\ }\textbf
  {\bibinfo {volume} {93}},\ \bibinfo {pages} {176402} (\bibinfo {year}
  {2004})}\BibitemShut {NoStop}%
\bibitem [{\citenamefont {Huertas-Hernando}\ \emph {et~al.}(2006)\citenamefont
  {Huertas-Hernando}, \citenamefont {Guinea},\ and\ \citenamefont
  {Brataas}}]{curvature2}%
  \BibitemOpen
  \bibfield  {author} {\bibinfo {author} {\bibfnamefont {D.}~\bibnamefont
  {Huertas-Hernando}}, \bibinfo {author} {\bibfnamefont {F.}~\bibnamefont
  {Guinea}}, \ and\ \bibinfo {author} {\bibfnamefont {A.}~\bibnamefont
  {Brataas}},\ }\href {\doibase 10.1103/PhysRevB.74.155426} {\bibfield
  {journal} {\bibinfo  {journal} {Phys. Rev. B}\ }\textbf {\bibinfo {volume}
  {74}},\ \bibinfo {pages} {155426} (\bibinfo {year} {2006})}\BibitemShut
  {NoStop}%
\bibitem [{\citenamefont {Hou}\ \emph {et~al.}(2007)\citenamefont {Hou},
  \citenamefont {Chamon},\ and\ \citenamefont {Mudry}}]{cham1}%
  \BibitemOpen
  \bibfield  {author} {\bibinfo {author} {\bibfnamefont {C.-Y.}\ \bibnamefont
  {Hou}}, \bibinfo {author} {\bibfnamefont {C.}~\bibnamefont {Chamon}}, \ and\
  \bibinfo {author} {\bibfnamefont {C.}~\bibnamefont {Mudry}},\ }\href
  {\doibase 10.1103/PhysRevLett.98.186809} {\bibfield  {journal} {\bibinfo
  {journal} {Phys. Rev. Lett.}\ }\textbf {\bibinfo {volume} {98}},\ \bibinfo
  {pages} {186809} (\bibinfo {year} {2007})}\BibitemShut {NoStop}%
\bibitem [{\citenamefont {Cheianov}\ \emph {et~al.}(2009)\citenamefont
  {Cheianov}, \citenamefont {Fal'ko}, \citenamefont {Syljuasen},\ and\
  \citenamefont {Altshuler}}]{keku}%
  \BibitemOpen
  \bibfield  {author} {\bibinfo {author} {\bibfnamefont {V.}~\bibnamefont
  {Cheianov}}, \bibinfo {author} {\bibfnamefont {V.}~\bibnamefont {Fal'ko}},
  \bibinfo {author} {\bibfnamefont {O.}~\bibnamefont {Syljuasen}}, \ and\
  \bibinfo {author} {\bibfnamefont {B.}~\bibnamefont {Altshuler}},\ }\href@noop
  {} {\bibfield  {journal} {\bibinfo  {journal} {Solid State Comm.}\ }\textbf
  {\bibinfo {volume} {149}},\ \bibinfo {pages} {1499 } (\bibinfo {year}
  {2009})}\BibitemShut {NoStop}%
\bibitem [{\citenamefont {Jeong}\ \emph {et~al.}(2011)\citenamefont {Jeong},
  \citenamefont {Shin},\ and\ \citenamefont {Lee}}]{curvature3}%
  \BibitemOpen
  \bibfield  {author} {\bibinfo {author} {\bibfnamefont {J.-S.}\ \bibnamefont
  {Jeong}}, \bibinfo {author} {\bibfnamefont {J.}~\bibnamefont {Shin}}, \ and\
  \bibinfo {author} {\bibfnamefont {H.-W.}\ \bibnamefont {Lee}},\ }\href
  {\doibase 10.1103/PhysRevB.84.195457} {\bibfield  {journal} {\bibinfo
  {journal} {Phys. Rev. B}\ }\textbf {\bibinfo {volume} {84}},\ \bibinfo
  {pages} {195457} (\bibinfo {year} {2011})}\BibitemShut {NoStop}%
\bibitem [{\citenamefont {Iadecola}\ \emph {et~al.}(2013)\citenamefont
  {Iadecola}, \citenamefont {Campbell}, \citenamefont {Chamon}, \citenamefont
  {Hou}, \citenamefont {Jackiw}, \citenamefont {Pi},\ and\ \citenamefont
  {Kusminskiy}}]{cham3}%
  \BibitemOpen
  \bibfield  {author} {\bibinfo {author} {\bibfnamefont {T.}~\bibnamefont
  {Iadecola}}, \bibinfo {author} {\bibfnamefont {D.}~\bibnamefont {Campbell}},
  \bibinfo {author} {\bibfnamefont {C.}~\bibnamefont {Chamon}}, \bibinfo
  {author} {\bibfnamefont {C.-Y.}\ \bibnamefont {Hou}}, \bibinfo {author}
  {\bibfnamefont {R.}~\bibnamefont {Jackiw}}, \bibinfo {author} {\bibfnamefont
  {S.-Y.}\ \bibnamefont {Pi}}, \ and\ \bibinfo {author} {\bibfnamefont {S.~V.}\
  \bibnamefont {Kusminskiy}},\ }\href {\doibase 10.1103/PhysRevLett.110.176603}
  {\bibfield  {journal} {\bibinfo  {journal} {Phys. Rev. Lett.}\ }\textbf
  {\bibinfo {volume} {110}},\ \bibinfo {pages} {176603} (\bibinfo {year}
  {2013})}\BibitemShut {NoStop}%
\bibitem [{\citenamefont {Iadecola}\ \emph {et~al.}(2014)\citenamefont
  {Iadecola}, \citenamefont {Neupert},\ and\ \citenamefont {Chamon}}]{cham2}%
  \BibitemOpen
  \bibfield  {author} {\bibinfo {author} {\bibfnamefont {T.}~\bibnamefont
  {Iadecola}}, \bibinfo {author} {\bibfnamefont {T.}~\bibnamefont {Neupert}}, \
  and\ \bibinfo {author} {\bibfnamefont {C.}~\bibnamefont {Chamon}},\ }\href
  {\doibase 10.1103/PhysRevB.89.115425} {\bibfield  {journal} {\bibinfo
  {journal} {Phys. Rev. B}\ }\textbf {\bibinfo {volume} {89}},\ \bibinfo
  {pages} {115425} (\bibinfo {year} {2014})}\BibitemShut {NoStop}%
\bibitem [{\citenamefont {Katsnelson}\ \emph {et~al.}(2006)\citenamefont
  {Katsnelson}, \citenamefont {Novoselov},\ and\ \citenamefont {Geim}}]{Klein}%
  \BibitemOpen
  \bibfield  {author} {\bibinfo {author} {\bibfnamefont {M.~I.}\ \bibnamefont
  {Katsnelson}}, \bibinfo {author} {\bibfnamefont {K.~S.}\ \bibnamefont
  {Novoselov}}, \ and\ \bibinfo {author} {\bibfnamefont {A.~K.}\ \bibnamefont
  {Geim}},\ }\href {\doibase 10.1038/nphys384} {\bibfield  {journal} {\bibinfo
  {journal} {Nat. Phys.}\ }\textbf {\bibinfo {volume} {2}},\ \bibinfo {pages}
  {620} (\bibinfo {year} {2006})}\BibitemShut {NoStop}%
\bibitem [{\citenamefont {Tikhonenko}\ \emph {et~al.}(2009)\citenamefont
  {Tikhonenko}, \citenamefont {Kozikov}, \citenamefont {Savchenko},\ and\
  \citenamefont {Gorbachev}}]{disord8}%
  \BibitemOpen
  \bibfield  {author} {\bibinfo {author} {\bibfnamefont {F.~V.}\ \bibnamefont
  {Tikhonenko}}, \bibinfo {author} {\bibfnamefont {A.~A.}\ \bibnamefont
  {Kozikov}}, \bibinfo {author} {\bibfnamefont {A.~K.}\ \bibnamefont
  {Savchenko}}, \ and\ \bibinfo {author} {\bibfnamefont {R.~V.}\ \bibnamefont
  {Gorbachev}},\ }\href {\doibase 10.1103/PhysRevLett.103.226801} {\bibfield
  {journal} {\bibinfo  {journal} {Phys. Rev. Lett.}\ }\textbf {\bibinfo
  {volume} {103}},\ \bibinfo {pages} {226801} (\bibinfo {year}
  {2009})}\BibitemShut {NoStop}%
\bibitem [{\citenamefont {Aleiner}\ and\ \citenamefont
  {Efetov}(2006)}]{disord1}%
  \BibitemOpen
  \bibfield  {author} {\bibinfo {author} {\bibfnamefont {I.~L.}\ \bibnamefont
  {Aleiner}}\ and\ \bibinfo {author} {\bibfnamefont {K.~B.}\ \bibnamefont
  {Efetov}},\ }\href@noop {} {\bibfield  {journal} {\bibinfo  {journal} {Phys.
  Rev. Lett.}\ }\textbf {\bibinfo {volume} {97}},\ \bibinfo {pages} {236801}
  (\bibinfo {year} {2006})}\BibitemShut {NoStop}%
\bibitem [{\citenamefont {Morpurgo}\ and\ \citenamefont
  {Guinea}(2006)}]{disord2}%
  \BibitemOpen
  \bibfield  {author} {\bibinfo {author} {\bibfnamefont {A.~F.}\ \bibnamefont
  {Morpurgo}}\ and\ \bibinfo {author} {\bibfnamefont {F.}~\bibnamefont
  {Guinea}},\ }\href {\doibase 10.1103/PhysRevLett.97.196804} {\bibfield
  {journal} {\bibinfo  {journal} {Phys. Rev. Lett.}\ }\textbf {\bibinfo
  {volume} {97}},\ \bibinfo {pages} {196804} (\bibinfo {year}
  {2006})}\BibitemShut {NoStop}%
\bibitem [{\citenamefont {McCann}\ \emph {et~al.}(2006)\citenamefont {McCann},
  \citenamefont {Kechedzhi}, \citenamefont {Fal'ko}, \citenamefont {Suzuura},
  \citenamefont {Ando},\ and\ \citenamefont {Altshuler}}]{disord4}%
  \BibitemOpen
  \bibfield  {author} {\bibinfo {author} {\bibfnamefont {E.}~\bibnamefont
  {McCann}}, \bibinfo {author} {\bibfnamefont {K.}~\bibnamefont {Kechedzhi}},
  \bibinfo {author} {\bibfnamefont {V.~I.}\ \bibnamefont {Fal'ko}}, \bibinfo
  {author} {\bibfnamefont {H.}~\bibnamefont {Suzuura}}, \bibinfo {author}
  {\bibfnamefont {T.}~\bibnamefont {Ando}}, \ and\ \bibinfo {author}
  {\bibfnamefont {B.~L.}\ \bibnamefont {Altshuler}},\ }\href {\doibase
  10.1103/PhysRevLett.97.146805} {\bibfield  {journal} {\bibinfo  {journal}
  {Phys. Rev. Lett.}\ }\textbf {\bibinfo {volume} {97}},\ \bibinfo {pages}
  {146805} (\bibinfo {year} {2006})}\BibitemShut {NoStop}%
\bibitem [{\citenamefont {Ostrovsky}\ \emph {et~al.}(2006)\citenamefont
  {Ostrovsky}, \citenamefont {Gornyi},\ and\ \citenamefont {Mirlin}}]{disord6}%
  \BibitemOpen
  \bibfield  {author} {\bibinfo {author} {\bibfnamefont {P.~M.}\ \bibnamefont
  {Ostrovsky}}, \bibinfo {author} {\bibfnamefont {I.~V.}\ \bibnamefont
  {Gornyi}}, \ and\ \bibinfo {author} {\bibfnamefont {A.~D.}\ \bibnamefont
  {Mirlin}},\ }\href {\doibase 10.1103/PhysRevB.74.235443} {\bibfield
  {journal} {\bibinfo  {journal} {Phys. Rev. B}\ }\textbf {\bibinfo {volume}
  {74}},\ \bibinfo {pages} {235443} (\bibinfo {year} {2006})}\BibitemShut
  {NoStop}%
\bibitem [{\citenamefont {Ma\~nes}\ \emph {et~al.}(2007)\citenamefont
  {Ma\~nes}, \citenamefont {Guinea},\ and\ \citenamefont
  {Vozmediano}}]{disord3}%
  \BibitemOpen
  \bibfield  {author} {\bibinfo {author} {\bibfnamefont {J.~L.}\ \bibnamefont
  {Ma\~nes}}, \bibinfo {author} {\bibfnamefont {F.}~\bibnamefont {Guinea}}, \
  and\ \bibinfo {author} {\bibfnamefont {M.~A.~H.}\ \bibnamefont
  {Vozmediano}},\ }\href {\doibase 10.1103/PhysRevB.75.155424} {\bibfield
  {journal} {\bibinfo  {journal} {Phys. Rev. B}\ }\textbf {\bibinfo {volume}
  {75}},\ \bibinfo {pages} {155424} (\bibinfo {year} {2007})}\BibitemShut
  {NoStop}%
\bibitem [{\citenamefont {Hong}\ \emph {et~al.}(2009)\citenamefont {Hong},
  \citenamefont {Zou},\ and\ \citenamefont {Zhu}}]{scatteringtimes}%
  \BibitemOpen
  \bibfield  {author} {\bibinfo {author} {\bibfnamefont {X.}~\bibnamefont
  {Hong}}, \bibinfo {author} {\bibfnamefont {K.}~\bibnamefont {Zou}}, \ and\
  \bibinfo {author} {\bibfnamefont {J.}~\bibnamefont {Zhu}},\ }\href {\doibase
  10.1103/PhysRevB.80.241415} {\bibfield  {journal} {\bibinfo  {journal} {Phys.
  Rev. B}\ }\textbf {\bibinfo {volume} {80}},\ \bibinfo {pages} {241415}
  (\bibinfo {year} {2009})}\BibitemShut {NoStop}%
\bibitem [{\citenamefont {Monteverde}\ \emph {et~al.}(2010)\citenamefont
  {Monteverde}, \citenamefont {Ojeda-Aristizabal}, \citenamefont {Weil},
  \citenamefont {Bennaceur}, \citenamefont {Ferrier}, \citenamefont {Gu\'eron},
  \citenamefont {Glattli}, \citenamefont {Bouchiat}, \citenamefont {Fuchs},\
  and\ \citenamefont {Maslov}}]{helen}%
  \BibitemOpen
  \bibfield  {author} {\bibinfo {author} {\bibfnamefont {M.}~\bibnamefont
  {Monteverde}}, \bibinfo {author} {\bibfnamefont {C.}~\bibnamefont
  {Ojeda-Aristizabal}}, \bibinfo {author} {\bibfnamefont {R.}~\bibnamefont
  {Weil}}, \bibinfo {author} {\bibfnamefont {K.}~\bibnamefont {Bennaceur}},
  \bibinfo {author} {\bibfnamefont {M.}~\bibnamefont {Ferrier}}, \bibinfo
  {author} {\bibfnamefont {S.}~\bibnamefont {Gu\'eron}}, \bibinfo {author}
  {\bibfnamefont {C.}~\bibnamefont {Glattli}}, \bibinfo {author} {\bibfnamefont
  {H.}~\bibnamefont {Bouchiat}}, \bibinfo {author} {\bibfnamefont {J.~N.}\
  \bibnamefont {Fuchs}}, \ and\ \bibinfo {author} {\bibfnamefont {D.~L.}\
  \bibnamefont {Maslov}},\ }\href {\doibase 10.1103/PhysRevLett.104.126801}
  {\bibfield  {journal} {\bibinfo  {journal} {Phys. Rev. Lett.}\ }\textbf
  {\bibinfo {volume} {104}},\ \bibinfo {pages} {126801} (\bibinfo {year}
  {2010})}\BibitemShut {NoStop}%
\bibitem [{\citenamefont {Mucciolo}\ and\ \citenamefont
  {Lewenkopf}(2010)}]{disord7}%
  \BibitemOpen
  \bibfield  {author} {\bibinfo {author} {\bibfnamefont {E.~R.}\ \bibnamefont
  {Mucciolo}}\ and\ \bibinfo {author} {\bibfnamefont {C.~H.}\ \bibnamefont
  {Lewenkopf}},\ }\href {http://stacks.iop.org/0953-8984/22/i=27/a=273201}
  {\bibfield  {journal} {\bibinfo  {journal} {J. Phys.Condens. Matt.}\ }\textbf
  {\bibinfo {volume} {22}},\ \bibinfo {pages} {273201} (\bibinfo {year}
  {2010})}\BibitemShut {NoStop}%
\bibitem [{\citenamefont {McCann}\ and\ \citenamefont
  {Fal'ko}(2012)}]{disord5}%
  \BibitemOpen
  \bibfield  {author} {\bibinfo {author} {\bibfnamefont {E.}~\bibnamefont
  {McCann}}\ and\ \bibinfo {author} {\bibfnamefont {V.~I.}\ \bibnamefont
  {Fal'ko}},\ }\href {\doibase 10.1103/PhysRevLett.108.166606} {\bibfield
  {journal} {\bibinfo  {journal} {Phys. Rev. Lett.}\ }\textbf {\bibinfo
  {volume} {108}},\ \bibinfo {pages} {166606} (\bibinfo {year}
  {2012})}\BibitemShut {NoStop}%
\bibitem [{\citenamefont {de~Juan}(2013)}]{nonabelian}%
  \BibitemOpen
  \bibfield  {author} {\bibinfo {author} {\bibfnamefont {F.}~\bibnamefont
  {de~Juan}},\ }\href@noop {} {\bibfield  {journal} {\bibinfo  {journal} {Phys.
  Rev. B}\ }\textbf {\bibinfo {volume} {87}},\ \bibinfo {pages} {125419}
  (\bibinfo {year} {2013})}\BibitemShut {NoStop}%
\bibitem [{\citenamefont {Pachoud}\ \emph {et~al.}(2014)\citenamefont
  {Pachoud}, \citenamefont {Ferreira}, \citenamefont {\"Ozyilmaz},\ and\
  \citenamefont {Castro~Neto}}]{ferreira2}%
  \BibitemOpen
  \bibfield  {author} {\bibinfo {author} {\bibfnamefont {A.}~\bibnamefont
  {Pachoud}}, \bibinfo {author} {\bibfnamefont {A.}~\bibnamefont {Ferreira}},
  \bibinfo {author} {\bibfnamefont {B.}~\bibnamefont {\"Ozyilmaz}}, \ and\
  \bibinfo {author} {\bibfnamefont {A.~H.}\ \bibnamefont {Castro~Neto}},\
  }\href {\doibase 10.1103/PhysRevB.90.035444} {\bibfield  {journal} {\bibinfo
  {journal} {Phys. Rev. B}\ }\textbf {\bibinfo {volume} {90}},\ \bibinfo
  {pages} {035444} (\bibinfo {year} {2014})}\BibitemShut {NoStop}%
\bibitem [{\citenamefont {Rashba}(1960)}]{Rashaba}%
  \BibitemOpen
  \bibfield  {author} {\bibinfo {author} {\bibfnamefont {E.}~\bibnamefont
  {Rashba}},\ }\href@noop {} {\bibfield  {journal} {\bibinfo  {journal} {Sov.
  Phys. Solid. State.}\ }\textbf {\bibinfo {volume} {2}},\ \bibinfo {pages}
  {1109} (\bibinfo {year} {1960})}\BibitemShut {NoStop}%
\bibitem [{\citenamefont {Min}\ \emph {et~al.}(2006)\citenamefont {Min},
  \citenamefont {Hill}, \citenamefont {Sinitsyn}, \citenamefont {Sahu},
  \citenamefont {Kleinman},\ and\ \citenamefont {MacDonald}}]{intrinsic}%
  \BibitemOpen
  \bibfield  {author} {\bibinfo {author} {\bibfnamefont {H.}~\bibnamefont
  {Min}}, \bibinfo {author} {\bibfnamefont {J.~E.}\ \bibnamefont {Hill}},
  \bibinfo {author} {\bibfnamefont {N.~A.}\ \bibnamefont {Sinitsyn}}, \bibinfo
  {author} {\bibfnamefont {B.~R.}\ \bibnamefont {Sahu}}, \bibinfo {author}
  {\bibfnamefont {L.}~\bibnamefont {Kleinman}}, \ and\ \bibinfo {author}
  {\bibfnamefont {A.~H.}\ \bibnamefont {MacDonald}},\ }\href {\doibase
  10.1103/PhysRevB.74.165310} {\bibfield  {journal} {\bibinfo  {journal} {Phys.
  Rev. B}\ }\textbf {\bibinfo {volume} {74}},\ \bibinfo {pages} {165310}
  (\bibinfo {year} {2006})}\BibitemShut {NoStop}%
\bibitem [{\citenamefont {Weeks}\ \emph {et~al.}(2011)\citenamefont {Weeks},
  \citenamefont {Hu}, \citenamefont {Alicea}, \citenamefont {Franz},\ and\
  \citenamefont {Wu}}]{indium}%
  \BibitemOpen
  \bibfield  {author} {\bibinfo {author} {\bibfnamefont {C.}~\bibnamefont
  {Weeks}}, \bibinfo {author} {\bibfnamefont {J.}~\bibnamefont {Hu}}, \bibinfo
  {author} {\bibfnamefont {J.}~\bibnamefont {Alicea}}, \bibinfo {author}
  {\bibfnamefont {M.}~\bibnamefont {Franz}}, \ and\ \bibinfo {author}
  {\bibfnamefont {R.}~\bibnamefont {Wu}},\ }\href {\doibase
  10.1103/PhysRevX.1.021001} {\bibfield  {journal} {\bibinfo  {journal} {Phys.
  Rev. X}\ }\textbf {\bibinfo {volume} {1}},\ \bibinfo {pages} {021001}
  (\bibinfo {year} {2011})}\BibitemShut {NoStop}%
\bibitem [{\citenamefont {Jia}\ \emph {et~al.}(2015)\citenamefont {Jia},
  \citenamefont {Yan}, \citenamefont {Niu}, \citenamefont {Han}, \citenamefont
  {Zhu}, \citenamefont {Yu},\ and\ \citenamefont {Wu}}]{indiumexpt}%
  \BibitemOpen
  \bibfield  {author} {\bibinfo {author} {\bibfnamefont {Z.}~\bibnamefont
  {Jia}}, \bibinfo {author} {\bibfnamefont {B.}~\bibnamefont {Yan}}, \bibinfo
  {author} {\bibfnamefont {J.}~\bibnamefont {Niu}}, \bibinfo {author}
  {\bibfnamefont {Q.}~\bibnamefont {Han}}, \bibinfo {author} {\bibfnamefont
  {R.}~\bibnamefont {Zhu}}, \bibinfo {author} {\bibfnamefont {D.}~\bibnamefont
  {Yu}}, \ and\ \bibinfo {author} {\bibfnamefont {X.}~\bibnamefont {Wu}},\
  }\href {\doibase 10.1103/PhysRevB.91.085411} {\bibfield  {journal} {\bibinfo
  {journal} {Phys. Rev. B}\ }\textbf {\bibinfo {volume} {91}},\ \bibinfo
  {pages} {085411} (\bibinfo {year} {2015})}\BibitemShut {NoStop}%
\bibitem [{\citenamefont {Asmar}\ and\ \citenamefont
  {Ulloa}(2014)}]{isotropic}%
  \BibitemOpen
  \bibfield  {author} {\bibinfo {author} {\bibfnamefont {M.~M.}\ \bibnamefont
  {Asmar}}\ and\ \bibinfo {author} {\bibfnamefont {S.~E.}\ \bibnamefont
  {Ulloa}},\ }\href {\doibase 10.1103/PhysRevLett.112.136602} {\bibfield
  {journal} {\bibinfo  {journal} {Phys. Rev. Lett.}\ }\textbf {\bibinfo
  {volume} {112}},\ \bibinfo {pages} {136602} (\bibinfo {year}
  {2014})}\BibitemShut {NoStop}%
\bibitem [{\citenamefont {Huertas-Hernando}\ \emph {et~al.}(2009)\citenamefont
  {Huertas-Hernando}, \citenamefont {Guinea},\ and\ \citenamefont
  {Brataas}}]{mediateso}%
  \BibitemOpen
  \bibfield  {author} {\bibinfo {author} {\bibfnamefont {D.}~\bibnamefont
  {Huertas-Hernando}}, \bibinfo {author} {\bibfnamefont {F.}~\bibnamefont
  {Guinea}}, \ and\ \bibinfo {author} {\bibfnamefont {A.}~\bibnamefont
  {Brataas}},\ }\href {\doibase 10.1103/PhysRevLett.103.146801} {\bibfield
  {journal} {\bibinfo  {journal} {Phys. Rev. Lett.}\ }\textbf {\bibinfo
  {volume} {103}},\ \bibinfo {pages} {146801} (\bibinfo {year}
  {2009})}\BibitemShut {NoStop}%
\bibitem [{\citenamefont {Ochoa}\ \emph {et~al.}(2012)\citenamefont {Ochoa},
  \citenamefont {Castro~Neto},\ and\ \citenamefont {Guinea}}]{ellyot}%
  \BibitemOpen
  \bibfield  {author} {\bibinfo {author} {\bibfnamefont {H.}~\bibnamefont
  {Ochoa}}, \bibinfo {author} {\bibfnamefont {A.~H.}\ \bibnamefont
  {Castro~Neto}}, \ and\ \bibinfo {author} {\bibfnamefont {F.}~\bibnamefont
  {Guinea}},\ }\href {\doibase 10.1103/PhysRevLett.108.206808} {\bibfield
  {journal} {\bibinfo  {journal} {Phys. Rev. Lett.}\ }\textbf {\bibinfo
  {volume} {108}},\ \bibinfo {pages} {206808} (\bibinfo {year}
  {2012})}\BibitemShut {NoStop}%
\bibitem [{\citenamefont {Asmar}\ and\ \citenamefont {Ulloa}(2013)}]{biref}%
  \BibitemOpen
  \bibfield  {author} {\bibinfo {author} {\bibfnamefont {M.~M.}\ \bibnamefont
  {Asmar}}\ and\ \bibinfo {author} {\bibfnamefont {S.~E.}\ \bibnamefont
  {Ulloa}},\ }\href {\doibase 10.1103/PhysRevB.87.075420} {\bibfield  {journal}
  {\bibinfo  {journal} {Phys. Rev. B}\ }\textbf {\bibinfo {volume} {87}},\
  \bibinfo {pages} {075420} (\bibinfo {year} {2013})}\BibitemShut {NoStop}%
\bibitem [{\citenamefont {Katsnelson}\ \emph {et~al.}(2009)\citenamefont
  {Katsnelson}, \citenamefont {Guinea},\ and\ \citenamefont {Geim}}]{cluster}%
  \BibitemOpen
  \bibfield  {author} {\bibinfo {author} {\bibfnamefont {M.~I.}\ \bibnamefont
  {Katsnelson}}, \bibinfo {author} {\bibfnamefont {F.}~\bibnamefont {Guinea}},
  \ and\ \bibinfo {author} {\bibfnamefont {A.~K.}\ \bibnamefont {Geim}},\
  }\href {\doibase 10.1103/PhysRevB.79.195426} {\bibfield  {journal} {\bibinfo
  {journal} {Phys. Rev. B}\ }\textbf {\bibinfo {volume} {79}},\ \bibinfo
  {pages} {195426} (\bibinfo {year} {2009})}\BibitemShut {NoStop}%
\bibitem [{\citenamefont {Peskin}\ and\ \citenamefont
  {Schroeder}(1995)}]{peskin}%
  \BibitemOpen
  \bibfield  {author} {\bibinfo {author} {\bibfnamefont {M.~E.}\ \bibnamefont
  {Peskin}}\ and\ \bibinfo {author} {\bibfnamefont {D.~V.}\ \bibnamefont
  {Schroeder}},\ }\href@noop {} {\emph {\bibinfo {title} {An Introduction To
  Quantum Field Theory}}}\ (\bibinfo  {publisher} {Westview Press},\ \bibinfo
  {year} {1995})\BibitemShut {NoStop}%
\bibitem [{\citenamefont {Srednicki}(2007)}]{book}%
  \BibitemOpen
  \bibfield  {author} {\bibinfo {author} {\bibfnamefont {M.}~\bibnamefont
  {Srednicki}},\ }\href@noop {} {\emph {\bibinfo {title} {Quantum Field
  Theory}}}\ (\bibinfo  {publisher} {Cambridge University Press},\ \bibinfo
  {address} {New York},\ \bibinfo {year} {2007})\BibitemShut {NoStop}%
\bibitem [{\citenamefont {Ryu}\ \emph {et~al.}(2009)\citenamefont {Ryu},
  \citenamefont {Mudry}, \citenamefont {Hou},\ and\ \citenamefont
  {Chamon}}]{masses}%
  \BibitemOpen
  \bibfield  {author} {\bibinfo {author} {\bibfnamefont {S.}~\bibnamefont
  {Ryu}}, \bibinfo {author} {\bibfnamefont {C.}~\bibnamefont {Mudry}}, \bibinfo
  {author} {\bibfnamefont {C.-Y.}\ \bibnamefont {Hou}}, \ and\ \bibinfo
  {author} {\bibfnamefont {C.}~\bibnamefont {Chamon}},\ }\href {\doibase
  10.1103/PhysRevB.80.205319} {\bibfield  {journal} {\bibinfo  {journal} {Phys.
  Rev. B}\ }\textbf {\bibinfo {volume} {80}},\ \bibinfo {pages} {205319}
  (\bibinfo {year} {2009})}\BibitemShut {NoStop}%
\bibitem [{\citenamefont {Beenakker}(2008)}]{BeenakkerRMP}%
  \BibitemOpen
  \bibfield  {author} {\bibinfo {author} {\bibfnamefont {C.~W.~J.}\
  \bibnamefont {Beenakker}},\ }\href {\doibase 10.1103/RevModPhys.80.1337}
  {\bibfield  {journal} {\bibinfo  {journal} {Rev. Mod. Phys.}\ }\textbf
  {\bibinfo {volume} {80}},\ \bibinfo {pages} {1337} (\bibinfo {year}
  {2008})}\BibitemShut {NoStop}%
\bibitem [{\citenamefont {Novikov}(2007)}]{elastic}%
  \BibitemOpen
  \bibfield  {author} {\bibinfo {author} {\bibfnamefont {D.~S.}\ \bibnamefont
  {Novikov}},\ }\href {\doibase 10.1103/PhysRevB.76.245435} {\bibfield
  {journal} {\bibinfo  {journal} {Phys. Rev. B}\ }\textbf {\bibinfo {volume}
  {76}},\ \bibinfo {pages} {245435} (\bibinfo {year} {2007})}\BibitemShut
  {NoStop}%
\bibitem [{\citenamefont {Bardarson}\ \emph {et~al.}(2009)\citenamefont
  {Bardarson}, \citenamefont {Titov},\ and\ \citenamefont
  {Brouwer}}]{resonant2}%
  \BibitemOpen
  \bibfield  {author} {\bibinfo {author} {\bibfnamefont {J.~H.}\ \bibnamefont
  {Bardarson}}, \bibinfo {author} {\bibfnamefont {M.}~\bibnamefont {Titov}}, \
  and\ \bibinfo {author} {\bibfnamefont {P.~W.}\ \bibnamefont {Brouwer}},\
  }\href {\doibase 10.1103/PhysRevLett.102.226803} {\bibfield  {journal}
  {\bibinfo  {journal} {Phys. Rev. Lett.}\ }\textbf {\bibinfo {volume} {102}},\
  \bibinfo {pages} {226803} (\bibinfo {year} {2009})}\BibitemShut {NoStop}%
\bibitem [{\citenamefont {Xiao}\ \emph {et~al.}(2007)\citenamefont {Xiao},
  \citenamefont {Yao},\ and\ \citenamefont {Niu}}]{hallsttagered}%
  \BibitemOpen
  \bibfield  {author} {\bibinfo {author} {\bibfnamefont {D.}~\bibnamefont
  {Xiao}}, \bibinfo {author} {\bibfnamefont {W.}~\bibnamefont {Yao}}, \ and\
  \bibinfo {author} {\bibfnamefont {Q.}~\bibnamefont {Niu}},\ }\href {\doibase
  10.1103/PhysRevLett.99.236809} {\bibfield  {journal} {\bibinfo  {journal}
  {Phys. Rev. Lett.}\ }\textbf {\bibinfo {volume} {99}},\ \bibinfo {pages}
  {236809} (\bibinfo {year} {2007})}\BibitemShut {NoStop}%
\bibitem [{\citenamefont {De~Martino}\ \emph {et~al.}(2011)\citenamefont
  {De~Martino}, \citenamefont {H\"utten},\ and\ \citenamefont
  {Egger}}]{newsym}%
  \BibitemOpen
  \bibfield  {author} {\bibinfo {author} {\bibfnamefont {A.}~\bibnamefont
  {De~Martino}}, \bibinfo {author} {\bibfnamefont {A.}~\bibnamefont
  {H\"utten}}, \ and\ \bibinfo {author} {\bibfnamefont {R.}~\bibnamefont
  {Egger}},\ }\href {\doibase 10.1103/PhysRevB.84.155420} {\bibfield  {journal}
  {\bibinfo  {journal} {Phys. Rev. B}\ }\textbf {\bibinfo {volume} {84}},\
  \bibinfo {pages} {155420} (\bibinfo {year} {2011})}\BibitemShut {NoStop}%
\bibitem [{\citenamefont {Elias}\ \emph {et~al.}(2009)\citenamefont {Elias},
  \citenamefont {Nair}, \citenamefont {Mohiuddin}, \citenamefont {Morozov},
  \citenamefont {Blake}, \citenamefont {Halsall}, \citenamefont {Ferrari},
  \citenamefont {Boukhvalov}, \citenamefont {Katsnelson}, \citenamefont
  {Geim},\ and\ \citenamefont {Novoselov}}]{hydrogen}%
  \BibitemOpen
  \bibfield  {author} {\bibinfo {author} {\bibfnamefont {D.~C.}\ \bibnamefont
  {Elias}}, \bibinfo {author} {\bibfnamefont {R.~R.}\ \bibnamefont {Nair}},
  \bibinfo {author} {\bibfnamefont {T.~M.~G.}\ \bibnamefont {Mohiuddin}},
  \bibinfo {author} {\bibfnamefont {S.~V.}\ \bibnamefont {Morozov}}, \bibinfo
  {author} {\bibfnamefont {P.}~\bibnamefont {Blake}}, \bibinfo {author}
  {\bibfnamefont {M.~P.}\ \bibnamefont {Halsall}}, \bibinfo {author}
  {\bibfnamefont {A.~C.}\ \bibnamefont {Ferrari}}, \bibinfo {author}
  {\bibfnamefont {D.~W.}\ \bibnamefont {Boukhvalov}}, \bibinfo {author}
  {\bibfnamefont {M.~I.}\ \bibnamefont {Katsnelson}}, \bibinfo {author}
  {\bibfnamefont {A.~K.}\ \bibnamefont {Geim}}, \ and\ \bibinfo {author}
  {\bibfnamefont {K.~S.}\ \bibnamefont {Novoselov}},\ }\href {\doibase
  10.1126/science.1167130} {\bibfield  {journal} {\bibinfo  {journal}
  {Science}\ }\textbf {\bibinfo {volume} {323}},\ \bibinfo {pages} {610}
  (\bibinfo {year} {2009})}\BibitemShut {NoStop}%
\bibitem [{\citenamefont {Sakurai}(1993)}]{Sakurai}%
  \BibitemOpen
  \bibfield  {author} {\bibinfo {author} {\bibfnamefont {J.~J.}\ \bibnamefont
  {Sakurai}},\ }\href@noop {} {\emph {\bibinfo {title} {Modern Quantum
  Mechanics}}}\ (\bibinfo  {publisher} {Addison Wesley},\ \bibinfo {year}
  {1993})\BibitemShut {NoStop}%
\end{thebibliography}%

\end{document}